\documentclass[12pt]{article}
\usepackage{graphicx,epsfig, epsf}
\usepackage{amsmath, amssymb, hyperref}

\makeatletter
\renewcommand\section{\@startsection {section}{1}{\z@}%
                                   {-3.5ex \@plus -1ex \@minus -.2ex}
                                   {2.3ex \@plus.2ex}%
                                   {\normalfont\large\bfseries}}
\renewcommand\subsection{\@startsection{subsection}{2}{\z@}%
                                     {-3.25ex\@plus -1ex \@minus -.2ex}%
                                     {1.5ex \@plus .2ex}%
                                     {\normalfont\bfseries}}

\makeatother

\newcommand{\bea}{\begin{eqnarray}}
\newcommand{\eea}{\end{eqnarray}}
\newcommand{\be}{\begin{equation}}
\newcommand{\ee}{\end{equation}}

\newcommand{\noi}{\noindent}

\newcommand{\Tr}{\mathrm{Tr}}
\newcommand{\im}{\mathrm{Im}}
\newcommand{\re}{\mathrm{Re}}

\newcommand{\CC}{\mathcal{C}}
\newcommand{\CD}{\mathcal{D}}
\newcommand{\CH}{\mathcal{H}}
\newcommand{\CO}{\mathcal{O}}
\newcommand{\CQ}{\mathcal{Q}}

\newcommand{\half}{\frac{1}{2}}

\newcommand{\IC}{\mathbb{C}}

\newcommand{\IN}{\mathbb{N}}

\newcommand{\IQ}{\mathbb{Q}}

\newcommand{\IZ}{\mathbb{Z}}

\begin{document}
\begin{titlepage}

\begin{center}

\hfill ITFA-07-54 \\
\hfill RUNHETC-2007-30

\vskip 2 cm {\Large \bf A Modern Fareytail}
\vskip 1.25 cm
{Jan Manschot}\\
{\vskip 0.5cm  Institute for Theoretical Physics, University of Amsterdam\\
Valckenierstraat 65, 1018 XE Amsterdam, The Netherlands\\}
\vskip 0.75 cm
and
\vskip 0.75 cm
{Gregory W. Moore}\\
{\vskip 0.5cm  NHETC and Department of Physics and Astronomy, Rutgers University \\
Piscataway, NJ 08855-0849, USA}

\pagestyle{plain}
\end{center}

\vskip 2 cm

\begin{abstract}
\baselineskip=18pt \noi We revisit the ``fareytail expansions'' of
elliptic genera which have been used in discussions of the
AdS$_3$/CFT$_2$ correspondence and the OSV conjecture. We show how
to write such expansions without the use of the problematic
``fareytail transform.'' In particular, we show how to write a
general vector-valued modular form of non-positive weight as a
convergent sum over cosets of $SL(2,\mathbb{Z})$. This sum suggests
a new regularization of the gravity path integral in AdS$_3$,
resolves the puzzles associated with the ``fareytail transform,''
and leads to several new insights.  We discuss constraints on the
polar coefficients of negative weight modular forms arising from
modular invariance, showing how these are related to Fourier
coefficients of positive weight cusp forms. In addition, we discuss
the appearance of holomorphic anomalies in the context of the fareytail.
\end{abstract}
\end{titlepage}

\baselineskip=19pt

\tableofcontents

\section{Introduction}

The AdS/CFT correspondence  \cite{Maldacena:1997re,
Gubser:1998bc,Witten:1998qj,Aharony:1999ti} plays a central role in
string theory. While it has yet to be given a concise and precise
mathematical definition, it seems clear that part of the formulation
involves an equality of partition functions
 \be \label{eq:zisz}
\mathcal{Z}_{\mathrm{String}}=\mathcal{Z}_{\mathrm{CFT}} \ee
where $\mathcal{Z}_{\mathrm{String}}$  is the partition function of
a string theory (or M-theory) on a spacetime (or sum over
spacetimes) with asymptotics of the form $AdS_n \times K$, for a
compact space $K$, and $\mathcal{Z}_{\mathrm{CFT}}$ is the partition
function of a ``holographically dual'' conformal field theory
defined on the conformal boundary of $AdS_n$. The present paper
discusses partition functions in the context of AdS$_3$/CFT$_2$, in
which case Eq. (\ref{eq:zisz}) can be investigated with a high degree of
precision.

We consider Euclidean AdS$_3$ geometries whose conformal boundary
geometry is a torus. Thus, the partition functions depend on the
complex structure parameter $\tau$ of the torus. The Fourier
expansion of the partition function, given by
\be
\mathcal{Z}=\sum_{n=0}^\infty c(n)q^{n-\Delta},
\ee
with $q=e^{2\pi i \tau}$, contains a pole when $\mathrm{Im}(\tau)\to \infty$, corresponding to the
 light states with $n-\Delta<0$. The partition function is uniquely specified by the
polar degeneracies using holomorphy and modular invariance. The main
result of this paper is the description of a sum which completes the
polar terms to the full partition function $\mathcal{Z}$.  The sum is
roughly a sum of the polar terms over a coset of the modular group
$SL(2,\mathbb{Z})$, \footnote{In the following we will abbreviate
  $SL(2,\mathbb{Z})$ to $\Gamma$.} which is known as a Poincar\'e
series. One of the important novel insights of Ref. \cite{Dijkgraaf:2000fq}
is the connection between Poincar\'e series and sums over different
AdS$_3$ geometries with fixed asymptotic boundary conditions. This
led to the proposal that $\mathcal{Z}_{\mathrm{CFT}}$, written as a
Poincar\'e series, has the interpretation as a sum over partition
functions of string theory on different spacetimes with fixed
conformal boundary conditions. Such an expansion of $\mathcal{Z}_{\mathrm{CFT}}$ has acquired the name ``fareytail expansion'' in the
physics literature. \footnote{The name refers to the fact that the sum over $\Gamma_\infty\backslash
\Gamma$ may be identified with a sum over fractions $d/c$ in lowest
terms. These define Farey series. In the context of black hole state counting
the  terms with $c>1$ are  exponentially small and
thus represent the tail of the micro-canonical distribution of
states associated with the black hole geometry.}

A closer inspection shows that the naive Poincar\'e series for the
relevant partition functions are divergent and must be regularized.
Ref. \cite{Dijkgraaf:2000fq} proposed a certain regularization which
unfortunately does not equal $\mathcal{Z}_{\mathrm{CFT}}$, but
rather equals a related function. This function, $\mathcal{
  \tilde Z}_{\mathrm{CFT}}$, the   so-called
``fareytail transform''  of  $\mathcal{Z}_{\mathrm{CFT}}$, is of the
form   $\mathcal{
  \tilde Z}_{\mathrm{CFT}} = \CO \mathcal{Z}_{\mathrm{CFT}}$ where
  $\CO$ is a certain  pseudo-differential operator.
   Therefore, the Poincar\'e series could not be directly
interpreted as a confirmation of Eq. (\ref{eq:zisz}).

An important achievement of this paper is a regularized version of
the naive Poincar\'e series which is equal to
$\mathcal{Z}_\mathrm{CFT}$ and not $\mathcal{
  \tilde Z}_{\mathrm{CFT}}$.  Since we no longer need to
transform $\mathcal{Z}_\mathrm{CFT}$, we have obtained an
interpretation of $\mathcal{Z}_\mathrm{CFT}$ as a sum over partition
functions with fixed conformal boundary conditions. This new version
is therefore much more appealing from the point of view of the
AdS/CFT correspondence.

 This new regularization is an application of a beautiful paper by D.  Niebur
\cite{niebur1974}, following up on earlier work of Knopp
 \cite{knopp:1990}. Niebur's  regularization reduces to the one
 proposed by Ref. \cite{Denef:2007vg} in the context of the OSV conjecture \cite{ooguri:2004zv} for
 Calabi-Yau manifolds with $b_2(X)$ even, and to the one used in
 Ref. \cite{Manschot:2007zb} for the partition function of pure AdS$_3$ gravity. Historically, these methods
 go back to Rademacher's expression for the partition function  $p(n)$
 of integers as an infinite sum of Bessel functions (see
 Ref. \cite{apostol} for a  modern account) and to his work
 \cite{rademacher:1939} expressing  the modular invariant
 $j$-function as a sum over $\Gamma_\infty\backslash \Gamma$.

The new regularization is not only justified by stating that it is
more appealing from the point of view of AdS/CFT. It also solves
some fundamental problems related to the fareytail transform. These
problems recently came to light in the course of some discussions
initiated by Hirosi Ooguri, during which Don Zagier pointed out that
in fact $\mathcal{\tilde Z}_\mathrm{CFT}$ is not modular in general.
We give a simple explanation of this in section
\ref{subsec:fareyrevisited} below. Thus, the reliance on the
mathematical properties of the fareytail transform in Ref.
\cite{Dijkgraaf:2000fq}  was a mistake and is erroneous.
\footnote{In the case of  negative
  half-integer weight Jacobi forms, or negative integer weight
  vectors of modular forms the fareytail transform does preserve
  modularity. In the application to the OSV conjecture  used in
  Ref. \cite{Denef:2007vg} this is the reason the authors  restricted
  attention to Calabi-Yau manifolds $X$ with even $b_2(X)$.}
Section \ref{subsec:fareyrevisited} explains the problems of the
fareytail transform in more detail.

The fareytail transform has no strong support from physics either.
In particular, other studies of Eq. (\ref{eq:zisz}) did not confirm
the need for a modification to $\mathcal{
  \tilde Z}_{\mathrm{CFT}}$.  For example, the first terms of the
Fourier expansions in Eq. (\ref{eq:zisz}) match in the case of the
D1-D5 system without the need for the fareytail transform
\cite{deBoer:1998us, Maldacena:1999bp}. The fareytail expansions used in attempts
(Refs. \cite{Denef:2007vg, deBoer:2006vg, Gaiotto:2006ns}) to put
the OSV conjecture on a firm footing  require Eq. (\ref{eq:zisz})
without application of the fareytail transform.  More recently, the
study of pure gravity in AdS$_3$ did not indicate any need for a
fareytail transform \cite{Witten:2007kt, Maloney:2007ud}.  Finally, tests in
four-dimensions involving the singleton modes in $AdS_5/CFT_4$
supported Eq. (\ref{eq:zisz}) without the need for modification
\cite{Witten:1998wy,Belov:2004ht}.

Once we have regularized the naive Poincar\'e series, we have to
re-examine modular invariance. We find that in general the
regularized Poincar\'e series do not transform covariantly under
modular transformations. The partition functions still transform in
a controlled way, which can be made precise using so-called period
functions and Eichler cohomology.   Thus the choice of polar
degeneracies is not arbitrary, as discussed in depth in section
\ref{sec:anomalies}. Alternatively, one can obtain modular
invariance by addition of a suitable non-holomorphic term, as
discussed in section \ref{sec:nonholomorphic}.

The regularization does not spoil the semi-classical interpretation
of the Poincar\'e series. The modern fareytail is therefore well
suited for use in the original applications, in particular AdS/CFT
and phase transitions. In the context of the tests of the OSV
conjecture the modern fareytail does not invalidate the previous arguments in
the regime of strong topological string coupling, although it does
lead to further corrections in the problematic regime of weak
topological string coupling. In section  \ref{subsec:OSVconjecture}
we comment on the ``entropy enigma'' of \cite{Denef:2007vg},
showing, in the context of a toy model for the polar degeneracies,
how in the Rademacher expansion the extreme polar states give the
dominant contribution to degeneracies close to the cosmic censorship
bound.

\label{sec:motivation} In the remaining part of the introduction we
will review briefly the connection between Poincar\'e series and
sums over asymptotically AdS$_3$ geometries. Also the new
regularization will be motivated heuristically. The connection between
elements in $\Gamma$ and AdS$_3$ geometries was suggested in Ref.
\cite{Maldacena:1998bw} and refined somewhat in
Ref. \cite{Dijkgraaf:2000fq}. It is reviewed for example in Refs.
\cite{deBoer:2006vg, Moore:2004fg} and Ref. \cite{kraus-2007-0701} from a
supergravity perspective. Three-dimensional gravity has no local
degrees of freedom, so different geometries arise from different
global identifications.  Euclidean AdS$_3$ is topologically equal to
a solid (filled-in) torus. The asymptotic metric is given by
\be   ds^2  \sim  r^2 |d\phi + idt/l|^2+\frac{dr^2}{r^2} \ee \noi
for $r\to \infty$,  where $\phi$ and $t$ are respectively a spatial
angular coordinate and periodic time, $l$ is related to the
cosmological constant. $\phi$ and $t$ satisfy the periodicities
$\phi + it/l\sim \phi + it/l + n+m\tau$. A homology basis of the
torus is given by two primitive cycles $A$ and $B$ with unit
intersection $A\cap B=1$. We choose the $A$-cycle to be contractible
in case of the solid torus. A choice of $A$ determines the filling
of the torus and therefore the AdS$_3$ geometry. The choice
determines $B$ up to multiples of $A$ since $A\cap A=0$. The choice
of $A$ is made with respect to a distinguished homology basis
$\alpha$ and $\beta$, with $\alpha \cap \beta=1$. The periods of a
holomorphic one form $\omega$ over $\alpha$ and $\beta$ are given by
$\int_\alpha \omega=1$ and $\int_\beta\omega=\tau$. $A$ and $B$ are
integer linear combinations of $\alpha$ and $\beta$ preserving the
intersection number. This determines that the two oriented bases are
related by an element of $\Gamma$. The complex structure parameter
of the torus is then defined by
\be \tau'=\frac{\int_B\omega}{\int_A
\omega}=\frac{a\tau +b}{c\tau+d}, \qquad \left(\begin{array}{cc} a &
b
  \\ c & d\end{array}\right)\in   \Gamma.
\ee
Since a choice of $A$ determines
$B$ only up to a multiple of $A$ we find that AdS$_3$ geometries are
related to elements of $\Gamma_\infty\backslash \Gamma$.
$\Gamma_\infty$ is the parabolic subgroup of $\Gamma$ of elements
$$\begin{pmatrix} 1 & n   \\ 0 & 1
  \\ \end{pmatrix} $$ for $n \in \mathbb{Z}.$
\noi
Note that the $\tau'$'s correspond to equivalent asymptotic tori, but
that they represent different fillings of the tori. We can see
what different choices of $A$ correspond to in gravity. For example
when the primitive contractible cycle is $\Delta(\phi + it)\sim 1$,
the spatial circle is contractible and we have periodic time, this is
thermal AdS$_3$. In case we take $\Delta(\phi + it)\sim \tau$, the
spatial circle is non-contractible and thus we have a black hole
geometry, this is the BTZ black hole \cite{Banados:1992wn}. The
Einstein-Hilbert action can be renormalized to obtain a finite answer
\cite{Henningson:1998gx, Skenderis:2002wp}. We find for the action of both geometries
\be
S_\mathrm{thermal}=-\frac{2\pi i}{24}\left(c_L \tau - c_R \bar \tau
\right), \qquad S_\mathrm{BTZ}=-\frac{2\pi
  i}{24}\left(-\frac{c_L}{\tau} +\frac{c_R}{\bar \tau} \right).
\ee
\noi where $c_L=c_R=\frac{3l}{2G}$. These actions naturally generalize
to actions of other geometries represented by $\Gamma_\infty\backslash \Gamma$. Eventually
we are interested in the description of supersymmetric geometries,
where the right moving part of the boundary SCFT is in the ground state. States
are therefore weighted by the exponent of the holomorphic part of the
action in the path integral. Such a holomorphic action can be
realized by adding an appropriate gravitational Chern-Simons term. Our
heuristic Ansatz for the gravity path integral is
\be \label{eq:gravpathintegral}
\mathcal{Z}_{\mathrm{grav}}(\tau)=\sum_{\Gamma_\infty \backslash \Gamma}
e^{-\frac{2\pi i c_L}{24}(\frac{a\tau+b}{c\tau +d})} \ee

\noi This sum is already similar to one of the main results of this
paper, Eq. (\ref{eq:sumcd}). The partition function is   not
convergent, so a suitable regularization is necessary. We will
determine the divergence and subtract that from the path integral.
We can rewrite the exponent for $c\neq 0$ as \be e^{-2\pi i
  \left(\frac{c_L}{24}\frac{a}{c}-\frac{\frac{c_L}{24}}{c(c\tau+d)}\right)}=e^{\left(-2\pi i \frac{c_L}{24}\frac{a}{c}\right)}\left(\sum_{l=0}^\infty
\frac{\left(2\pi i\frac{\frac{c_L}{24}}{c(c\tau+d)}\right)^{l}}{l!}\right).
\ee

\noi Convergence of the sum over $(c,d)$ can be shown for all but
the term with $l=0$. We thus have to subtract the term with $l=0$ from the
sum. We arrive at
\be
\label{eq:reggravpath}
\mathcal{Z}_{\mathrm{grav}}(\tau)=\sum_{\Gamma_\infty \backslash \Gamma}
e^{-\frac{2\pi i c_L}{24}(\frac{a\tau+b}{c\tau +d})}-r(a,c), \qquad
r(a,c)=\left\{ \begin{array}{cc}e^{-\frac{2\pi i c_L}{24}\frac{a}{c}},
  & c\neq 0, \\ 0, & c=0. \end{array} \right.
\ee
This is the regularization suggested in Ref. \cite{Manschot:2007zb}
for the partition function of pure gravity in AdS$_3$. In case of
negative integer weight more terms need to be subtracted. This was
proposed earlier in
  Ref. \cite{Denef:2007vg}. Equations (\ref{eq:tdef}) to (\ref{Regtrm})
  explain   a very natural
generalization of this idea to non-integer weight. We propose that
this is the proper way to regularize the gravity path integral in
AdS$_3$ because in contrast to the fareytail transform the degeneracies
are not changed with respect to the CFT partition function and it
holds for general weights depending on the matter content of the
theory.

As indicated earlier, our main interest lies in the study of
 supergravity in AdS$_3$ with a supersymmetric boundary
 theory. Ref. \cite{Dijkgraaf:2000fq} considered the  case of type II
 string  theories on $AdS_3 \times K$ whose holographic  dual is an
 $\mathcal{N}=(4,4)$ superconformal field theory. A second application
 is to the AdS$_3$ supergravities with  $(0,4)$ supersymmetry. These
 arise in the context of M-theory black holes. The relevant partition
 function of the SCFT is the so-called elliptic genus. This is an index \footnote{It is the
 character-valued index of the right-moving Dirac-Ramond operator.}
 and therefore one might hope to find an exact semi-classical
 expansion of these functions. This gives some motivation for
 expecting a fareytail expansion.

We denote the elliptic genus by  $\chi(\tau, z)$, where $z$ is a
vector in a complex vector space. Standard properties of
superconformal field theory show that $\chi(\tau,z)$ transforms as a
(generalized) Jacobi form. In the case of the $\mathcal{N}=(2,2)$
elliptic genus, $z$ is one dimensional. The dependence on $z$ arises
from the presence of gauge fields in the bulk of AdS$_3$. Applying
the reasoning as before, we expect an expansion of the form
\be
\label{eq:formalps}
\chi(\tau, z) \sim \sum_{\Gamma_\infty\backslash
  \Gamma} \chi^{-}\left(\frac{a\tau+b}{c\tau+d},\frac{z}{c\tau +d}\right) .
\ee
$\chi^{-}(\tau,z)$ is a truncation of the Fourier expansion of
$\chi(\tau,z)$. This truncation corresponds to states which are not
sufficiently massive to form black holes. The partition function
$\chi(\tau,z)$ written as in Eq. (\ref{eq:formalps}), is a sum of the
light excitations over all the geometries given by $\Gamma_\infty \backslash \Gamma$.
 Section \ref{sec:fareytail} presents the mathematically rigorous
 fareytail for the elliptic genera. We refer to Section \ref{sec:applications}
for more details on the physical interpretation and the special role
played by the constant term in the Fourier expansion.

We conclude the introduction by giving the outline of the paper. In
Section \ref{sec:review} we review relevant aspects of partition
functions in CFT's. Section \ref{sec:fareytail} presents the modern
fareytail, including the expressions for elliptic genera, relevant
for the D1-D5 systems and $\mathcal{N}=2$ black holes. The
derivations are relegated to the Appendix. Section
\ref{sec:anomalies} discusses possible modular anomalies arising
from the regularization, together with the constraints imposed on
the polar terms. We discuss applications of the fareytail
expansion in Section \ref{sec:applications} and indicate novel
aspects of the modern fareytail. Section \ref{sec:nonholomorphic}
discusses potential holomorphic anomalies in the  partition
functions. We finish with some concluding remarks in Section
\ref{sec:conclusion}.

\section{Modular Invariance and Elliptic Genera}
\label{sec:review}\setcounter{equation}{0} We review very briefly
 invariance under $\Gamma$ of conformal field theory partition
functions on a torus, and point out aspects which are important for
our discussion. A torus is conveniently represented as the quotient
of the complex plane by a lattice $\Lambda$, spanned by generators
$\vec \alpha$ and $\vec \beta$. A conformal field theory on a torus
does not depend on the size of the torus nor on any absolute
direction of the lattice vectors, so it naturally depends only on
$\tau=\left(\vec \alpha \cdot \vec \beta+i |\vec \alpha \times \vec
\beta|\right)/|\vec \alpha|^2$. The theory should furthermore be
invariant under large orientation preserving reparametrizations
 which leave the lattice invariant. This is the famous group
$\Gamma=\left\{ \left(\begin{array}{cc} a & b\\ c & d
\end{array}\right): a,b,c,d \in \mathbb{Z},\,ad-bc=1\right\}$.

The partition function of a bosonic conformal field theory on a torus is defined by
\be
\mathcal{Z}(\tau)=\Tr \left(q^{L_0-\frac{c_L}{24}} \bar q^{\bar L_0-\frac{c_R}{24}}\right).
\ee

\noi A factor of $(-1)^F$ must be included depending on the boundary
conditions (Neveu-Schwarz or Ramond) when a partition function with
fermions is considered. $\mathcal{Z}(\tau)$ must be regular in the
upper half plane $\mathcal{H}: \im(\tau)>0$. Possible poles occur only
at $i\infty \cup \mathbb{Q}$. Modular invariance has important consequences for the content of
holomorphic and anti-holomorphic sectors.

Elliptic genera are distinguished partition functions of
supersymmetric CFT's because they contain important topological
information. We briefly review now elliptic genera in
$\mathcal{N}=(4,4)$ and $(0,4)$ SCFT's. Both appear as boundary
conformal field theory of certain supergravities in AdS$_3$.
$\mathcal{N}=(4,4)$ SCFT's arise in the context of D1-D5 systems,
the SCFT is a sigma model with target space Sym$^{m}(X)$ at the
orbifold point in moduli space \cite{Strominger:1996sh}. $X$ is a
two complex dimensional Ricci flat manifold. $\mathcal{N}=(0,4)$
SCFT's arise in the study of four dimensional $\mathcal{N}=2$ black
holes, which can be described by wrapped M5 branes with fluxes after
an uplift to M-theory \cite{Maldacena:1997de}.

We use $\mathcal{N}=(2,2)$ notation to calculate the elliptic genus
of $\mathcal{N}=(4,4)$ SCFT. The elliptic genus of an
$\mathcal{N}=(2,2)$ SCFT is defined as a trace over the
Ramond-Ramond sector by
 \be \chi(\tau,z)_X =\Tr_\mathrm{RR} (-)^F
y^{J_0} q^{L_0-\frac{c_L}{24}} \bar q^{\bar L_0-\frac{c_R}{24}}.
\ee

\noi $F$ is the fermion number and given by $\frac{1}{2}(J_0-\tilde J_0)$.
$\chi(\tau,z)_X$ is independent of $\bar q$, because the insertion
of $(-)^F$ projects to right moving ground states. When the SCFT is
a sigma model, the elliptic genus can be shown to equal an integral
of a Chern character times the Todd class over $X$. This point of
view leads to the following explicit expression for the elliptic
genus  \cite{Kawai:1993jk,gritsenko-991}
 \be \label{eq:intell}
\chi(\tau,z)_X=\int_X \prod_{i=1}^{d/2}\frac{\theta_1(\tau,
  z+\xi_i)}{\theta_1(\tau, \xi_i)}2\pi i \xi_i,
\ee

\noi where the $\xi_i$ are defined by
\be
c(T_X)=1+c_1(T_X)+\dots c_{d/2}(T_X) =\prod_{i=1}^{d/2}(1+2\pi i \xi_i).
\ee

\noi $\chi(\tau,z)_X$ reduces for different values of the parameter
$z$ to the Euler number, Hirzebruch signature or $\hat A$ genus.
$\theta_1(\tau, z)$ is the odd Jacobi theta function. For the
definition see   the appendix of Ref. \cite{Kawai:1993jk}.

The elliptic genus for a two complex dimensional K\"ahler
manifold $X$ with Euler number $\chi$ and Hirzebruch signature $\sigma$
can straightforwardly be calculated:
\be
\chi(\tau,z)_X=-\frac{\sigma}{16}\chi(\tau,z)_{K3}+\frac{3}{8\pi^2}(\sigma+\frac{2}{3}\chi)\frac{\left(\partial_z
  \theta_1(\tau,z)\right)^2}{\eta(\tau)^6},
\ee

\noi with
\be
\chi(\tau,
z)_{K3}=24\frac{\theta_3(\tau,z)^2}{\theta_3(\tau)^2}-2\frac{\theta_4(\tau)^4-\theta_2(\tau)^4}{\eta(\tau)^4}\frac{\theta_1(\tau,z)^2}{\eta(\tau)^2}.
\ee

\noi  Ref. \cite{Dijkgraaf:1996xw} explains how to write a generating
function for the elliptic genera of Sym$^{m}(X)$, starting from the
elliptic genus of $X$.

Transformation properties of the elliptic genus under $\Gamma$ can
be deduced from the CFT and as well from Eq. (\ref{eq:intell})
\cite{Kawai:1993jk}. Most important is the case when $c_1(T_X)=0$.
The elliptic genus transforms in this case as a weak Jacobi form of
weight $k=0$ and index $m=c_L/6=d/4$. Jacobi forms with weight $k$
and index $m$ transform in the following way \footnote{ Throughout
the paper we use the convention common in the math literature that
$e(x):=e^{2\pi i x}$. We will also frequently use the notation
$\gamma(\tau)=\frac{a\tau+b}{c\tau+d}$ and $j(\gamma,\tau) = c\tau +
d$ where $a,b,c,d$ are the familiar elements of $\gamma$ when
written as a $2\times 2 $  matrix. Warning: the use of
$j(\gamma,\tau)$ in the mathematics literature is not consistent, it
is also sometimes used to denote $(c\tau + d)^{\frac{1}{2}}$
multiplied with the appropriate unitary factor.}
\begin{eqnarray}
\label{eq:jactrans}
&\phi\left( \gamma(\tau),
\frac{z}{j(\gamma,\tau)}\right)=j(\gamma,\tau)^k e\left(\frac{m c z^2}{j(\gamma,\tau)}\right)\phi(\tau,z), \qquad \gamma=\left( \begin{array}{cc} a & b \\ c & d
\end{array} \right)\in \Gamma, \\
&\phi\left(\tau, z+\lambda \tau
+\mu\right)=(-1)^{2m(\lambda+\mu)}e\left(-m (\lambda^2 \tau + 2
\lambda z )\right)\phi(\tau,z), \qquad (\lambda, \,\, \mu) \in
\mathbb{Z}^2. \nonumber
\end{eqnarray}

\noi The transformation property in the second line follows from the
invariance of the SCFT under spectral flow. Spectral flow is a
symmetry of the algebra; the bosonic generators transform as
\be
L_n \to L_n+\lambda J_n+\frac{c}{6}\lambda^2\delta_{n,0}, \qquad
J_n\to J_n+\frac{c}{3}\lambda \delta_{n,0}.
\ee

\noi Integer spectral flow maps Ramond states to Ramond states and
Neveu-Schwartz to Neveu-Schwartz states, whereas half-integer spectral
flow exchanges the states in the two sectors. The elliptic genus does not transform
as a Jacobi form when $c_1(T_X)\neq 0$, but instead transforms with a shift.

We describe now some important properties of Jacobi forms. Proofs can
be found in  Ref. \cite{eichler}. We expand a weak Jacobi form as a Fourier
series
\be
\label{eq:fourierjacobi}
\phi(\tau, z)=\sum_{n\geq 0, l\in \mathbb{Z}}c(n,l) q^n y^l.
\ee

\noi The transformation property which is based on spectral flow
determines $c(n,l)$ to be a function only of $4mn-l^2$ and $l\mod
2m$. A Jacobi form is called a ``weak'' Jacobi form when $c(n,l)$ is
only non-zero when $4mn-l^2\geq -m^2$. Furthermore, we can deduce that $\phi(\tau, z)$ can be
decomposed into a vector-valued modular form and theta functions
\begin{equation}
\label{eq:thetadecomposition}
\phi(\tau, z)=\sum_{\mu \mod 2m} h_\mu(\tau) \theta_{m,\mu}(\tau,z),
\end{equation}

\noindent where $\mu$ is a coset representative
 $\mathbb{Z}/2m\mathbb{Z}$. $h_\mu(\tau, z)$ and
 $\theta_{m,\mu}(\tau,z)$ are given by
\begin{equation}
\label{eq:fourierjacobidecomp}
h_\mu(\tau)=\sum_{n=-\mu^2\mod 4m} c_\mu(n) q^{n/4m}, \qquad
\theta_{m,\mu}(\tau,z)=\sum_{{l\in \mathbb{Z} \atop l=\mu \mod 2m}}q^{l^2/4m}y^l,
\end{equation}

\noi with $c_\mu(n)=(-1)^{2ml}c(\frac{n+l^2}{4m},l),\,\,l=\mu \mod 2m$. All  the
information concerning the Fourier coefficients of $\phi(\tau,z)$ is
thus captured in $h_\mu(\tau)$. The theta functions transform as a modular vector under modular
transformations. The generators $S$ and $T$ of $\Gamma$ transform $\theta_{m,\mu}(\tau)$ to
\begin{eqnarray}
\label{eq:thetatransform1}
&\theta_{m,\mu}\left(\frac{-1}{\tau},\frac{z}{\tau}\right)&=\sqrt{\frac{\tau}{2mi}}
e\left(\frac{mz^2}{\tau}\right)\sum_{\nu \mod 2m}
e\left(-\frac{\mu \nu}{2m}\right) \theta_{m,\nu}(\tau,z), \\
&\theta_{m,\mu}(\tau+1,z)&=e\left(\frac{\mu^2}{4m}\right)\theta_{m,\mu}(\tau,z). \nonumber
\end{eqnarray}
For an unambiguous value of the square root,
we define $\log z$ to be given by $\log z := \log |z| + i\arg (z)$
with $-\pi < \arg (z) \leq \pi$. For general transformations under $\Gamma$, we define a matrix $M(\gamma)_{\nu}^\mu$ by
\be
\theta_{m,\mu}\left(\gamma(\tau),\frac{z}{c\tau +d}\right)=j(\gamma,\tau)^{\frac{1}{2}}e\left(\frac{mcz^2}{c\tau+d}\right)M^{-1}(\gamma)_{\nu}^{\mu}\theta_{m,\nu}(\tau,z)
\ee
Such that we have for $h_\mu(\gamma(\tau))$ by Eq. (\ref{eq:jactrans})
\be
h_\mu(\gamma(\tau))=j(\gamma,\tau)^{k-\frac{1}{2}}M(\gamma)_\mu^\nu h_\nu(\tau),
\ee
The introduction of $M(\gamma)_\mu^\nu$ is convenient for a generalization to similar
partition functions, as for example elliptic genera of $\mathcal{N}=(0,4)$ SCFT's.

We will very briefly review the elliptic genera of
$\mathcal{N}=(0,4)$ SCFT arising in the study of $\mathcal{N}=2$
M-theory black holes. We refer to the references
\cite{Maldacena:1997de,Minasian:1999qn,
  deBoer:2006vg, Gaiotto:2006wm,kraus-2007-0701} for the precise
details. Ref. \cite{Denef:2007vg} performs a similar analysis which
results in the same partition function from the point of view of IIA
string theory. Elliptic genera in an $\mathcal{N}=(0,4)$ SCFT are
defined in a similar manner to those in $\mathcal{N}=(2,2)$ SCFT.
However, we need to insert a factor of $F^2$ in order to obtain a
non-zero answer, because of the cancellation between bosonic and
fermionic degrees of freedom on the supersymmetric side of the
$\mathcal{N}=(0,4)$ SCFT. This sum projects on half-BPS states on
the supersymmetric side. The CFT arises after reducing the degrees
of freedom from an M5-brane with world volume $\Sigma\times T^2$ to
$T^2$ where  $\Sigma$ is an ample divisor Poincar\'e dual to $P\in
H^2(X,\IZ)$ in a  Calabi-Yau threefold $X$
\cite{Maldacena:1997de,Minasian:1999qn}. We will often write $P$ in
place of $\Sigma$ for quantities that only depend on the homology
class of $\Sigma$.  The $\mathcal{N}=(0,4)$ elliptic genus of this
SCFT is given by Ref. \cite{deBoer:2006vg, Gaiotto:2006wm}
\begin{eqnarray}\label{eq:04eg}
\chi(\tau, z)_{P}&=&\Tr_R\left[ \frac{1}{2}F^2 (-)^F e(P \cdot Q/2) \right. \\
&&\times \left. e\left(\tau \left(L_0-\frac{c_L}{24} \right)-\bar
\tau \left(\bar L_0-\frac{c_R}{24}\right) + z \cdot Q \right)
\right], \nonumber
\end{eqnarray}

\noi where $Q\in H^4(X;\mathbb{Z}) $ are M2 brane charges of the
black hole, (generated by fluxes on the M5 brane) and $z\in
H^2(X;\mathbb{C})$.

A spectral flow exists in this SCFT similar to the spectral flow in
$\mathcal{N}=(2,2)$ SCFT allowing one to give an analogous
``singleton'' decomposition in terms of theta functions. In order to
write this out we need to introduce some notation. The lattice
$L_X:=\iota_P^*(H^2(X;\mathbb{Z}))\subset H^2(P;\mathbb{Z})$ has
signature $(+^1,-^{b_2-1})$ where $b_2=\dim H_2(X)$. The integral
quadratic form on $L_X$ can be written in terms of the intersection
numbers $d_{abc}$  of $X$ by introducing an integral basis $D_a$ for
$H_4(X,\mathbb{Z})$ and writing $v^2=d_{abc}P^a v^b v^c$. The
sublattice $L_X \oplus L_X^\perp \subset H^2(P,\IZ)$ is of index
$\det D_{ab}$ where $D_{ab}:=d_{abc}P^c$. We choose a set of glue
vectors, $\mu$, i.e. a rule for lifting   elements of the
discriminant group $[\mu]\in\CD=H^2(P,\IZ)/(L_X \oplus L_X^\perp)$
to $\mu \in H^2(P,\IZ)$ so that any vector $v\in H^2(P,\IZ)$ can be
written $v = v^{\|} + \mu + v^{\bot}$, with $v^{\|}\in L_X,
v^{\bot}\in L_X^{\bot}$. Now $H^2(P;\IZ)\otimes \IQ$ has a
projection to the negative and positive definite subspaces and we
denote this projection by $v \to v_+ \oplus v_-$. If $X,Y\in
H^2(P;\IZ)\otimes \IQ$ and $f$ is holomorphic introduce the notation
\begin{equation} \label{shortnot}
 E[f(\tau) X \cdot Y] := e^{-2 \pi i \, f(\tau) \, X_- \cdot Y_-
- 2 \pi i \, f(\bar{\tau}) \, X_+ \cdot Y_+}, \qquad E[A+B] :=
E[A]\,E[B].
\end{equation}
We now introduce the Siegel-Narain theta function for the lattice
$L_X$:
\begin{equation}
\label{eq:siegelnarain}
 \Theta_\mu(\tau,z) := \sum_{v\in L_X } E\left[\frac{\tau}{2}
 \left(\frac{P}{2} + \mu^\| + v\right)^2 +
 \left(\frac{P}{2} + \mu^\|+ v \right) \cdot \left(z + \frac{P}{2}\right)\right],
\end{equation}
where $z \in L_X \otimes \IC$ and the projection to $(z_+,z_-)$ is
extended $\IC$-linearly. Note that $\Theta_\mu$ is non-holomorphic
in $\tau$. In terms of these theta functions we have the
decomposition:
\be\label{singlem5} \chi(\tau,  z)_{P}=\sum_{\mu }
  h_\mu(\tau) \Theta_\mu(\tau,z),
\ee
Here  the functions $h_\mu(\tau)$ are holomorphic in $\tau$ and have
no singularities in the upper half plane.

Modular transformations act on the argument of the theta function
according to:
\begin{equation}\label{eq:modacttz}
\gamma\cdot (\tau,z_+, z_-): = \left(\frac{a \tau + b}{c \tau + d},
\frac{z_+}{c\bar\tau +d}, \frac{z_-}{c \tau +d} \right).
\end{equation}
 We will abbreviate (\ref{eq:modacttz})  as
$\gamma\cdot(\tau,z)$. Now, for generic $SU(3)$ holonomy Calabi-Yau,
duality symmetries in string theory imply:
\begin{equation} \label{SdualityZ}
 \chi(\gamma \cdot(\tau,z)) = \tilde M(\gamma) \, (c\tau +d)^{-3/2}
 (c\bar{\tau}+d)^{1/2} \, E\left[\frac{c}{c \tau+d} \frac{z^2}{2}\right] \,
\chi(\tau,z),
\end{equation}
where $\tilde M$ is a multiplier system given in Ref.
\cite{Denef:2007vg}. From this one deduces that the vector of
modular forms $h_\mu(\tau)$ transforms with weight
$\frac{-b_2}{2}-1$. These functions have a Fourier expansion
\be h_\mu(\tau)= \sum_{n\geq 0} H_\mu(n) e((n-\Delta_\mu)\tau), \ee
where
\be \Delta_\mu = \frac{c_L}{24} + {\rm Max}_{v \in L_X^{\bot}} \half
(v + \mu^{\bot})^2,  \ee
and $c_L=\chi(P) = P^3 + c_2(X)\cdot P$ is the Euler character of a
generic smooth divisor in the linear system $\vert P\vert$. (In
taking the maximum note that the quadratic form on $L_X^{\bot}$ is
negative definite.)  For $\mu=0$ the leading coefficient
$H_{\mu=0}(0)= (-1)^{I_P-1}I_P$ where $I_P=\frac{P^3}{6} +
\frac{c_2(X)\cdot P}{12}$ is the Euler character of the linear
system $\vert P \vert$.

There is also a supergravity viewpoint on the decomposition   Eq.
(\ref{singlem5}). It can  be regarded as the singleton decomposition
of the M5-brane partition function. The general singleton
decomposition of the M5-brane partition function was given in Ref.
\cite{Moore:2004jv}, where it was explained that the discriminant
group $\CD$ is the group of Page charges in the presence of
$G$-flux.

Summarizing, we have seen the relevance of vector-valued modular
forms in the study of partition functions; the weight and
multiplier system are determined by the content and symmetries of
the theory.

\section{The Modern Fareytail}
\label{sec:fareytail}\setcounter{equation}{0} The previous section
introduced elliptic genera and some of their properties. It
motivated the study of vector-valued modular forms $f_\mu(\tau)$ of
non-positive weight $w$. This section describes a fareytail
expansion for vector-valued modular forms and subsequently for
elliptic genera. The novel aspect of our discussion is the absence
of the ``fareytail transform.''   A summary of the derivation of the
result is given in appendix \ref{sec:derivation}. Section
\ref{sec:anomalies} examines how the regularization preserves the
modular properties.

\subsection{Vector-valued modular forms}
\label{subsec:vector} This section states the fareytail expansion of
vector-valued modular forms in detail. Let us, then, consider a
vector-valued modular form  $f_\mu(\tau)$ transforming under
$\Gamma$, as
\be \label{eq:modtransform} f_\mu(\gamma(\tau))=j(\gamma,\tau)^w
M(\gamma)_\mu^\nu
 f_\nu(\tau), \ee
We will be concerned with forms of weight $w \leq 0$, where $w$ is
not necessarily integral. For example, for the elliptic genus
$w=-1/2$. For the OSV conjecture $w=-1-b_2/2$.  We therefore must
choose a branch of the log to define $j(\gamma,\tau)^w$ and we take
$\log z := \log \vert z \vert + i {\rm arg}(z)$ with $-\pi < {\rm
arg}(z) \leq \pi$. For the $(2,2)$ and $(0,4)$ elliptic genus the
multiplier system $M(\gamma)$ will turn out to be unitary matrices.
See appendix \ref{sec:multisystems}.

We assume $M(T)$ is diagonalizable, and hence $f_\mu$ has a Fourier
expansion
\be \label{fourexp}  f_\mu(\tau)=\sum_{m=0}^\infty
F_\mu(m)q^{m-\Delta_\mu}, \ee
\noi where $F_\mu(0)\not=0$. \footnote{In general we follow the
notation of Ref. \cite{Dijkgraaf:2000fq}. However, we have changed
the sign of $\Delta_\mu$ relative to this reference. Also, following
\cite{eichler} we denote the index of a Jacobi form by $m$, whereas
$k$ is used in Ref. \cite{Dijkgraaf:2000fq}. In this paper we use
$w$ for the weight of a vector-valued modular form; $k$ is the
weight of a Jacobi form.} Poles of $f_\mu(\tau)$ occur only at the
cusps, i.e. $\gamma(i\infty)$, $\gamma \in \Gamma$. The pole at
$\tau=i\infty$ arises from  the polar part $f^-(\tau)$ of the
partition function \be
f_\mu^-(\tau):=\sum_{m-\Delta_\mu<0}F_\mu(m)q^{m-\Delta_\mu}. \ee

\noi The Fourier coefficients $F_\mu(m)$ can be calculated by the
Rademacher circle method \cite{Dijkgraaf:2000fq, apostol,
rademacher1938:1}. Sufficient information to calculate them are the
Fourier coefficients $F_\mu(m)$ for $m-\Delta_\mu<0$, the weight
$w$, and the multiplier system. Starting from the Fourier
coefficients for general $m$, we can derive the fareytail expansion
of the partition function as a sum over the limit coset: $\lim_{K\to
\infty}(\Gamma_\infty\backslash \Gamma)_K=\lim_{K\to
\infty}\sum_{|c|\leq K}\sum_{{|d|\leq K} \atop
  {(c,d)=1}}$.

Some details are given in appendix \ref{sec:derivation}. The result
is a sum over the polar part
\begin{eqnarray}
\label{eq:sumcd}
&f_\mu(\tau)&=\frac{1}{2}F_\mu(\Delta_\mu) + \frac{1}{2}
\sum_{n-\Delta_\nu<0} \lim_{K\to \infty} \sum_{(\Gamma_\infty
  \backslash \Gamma)_K}  \\
&& j(\gamma,\tau)^{-w} M^{-1}(\gamma)_\mu^\nu F_\nu(n)
e((n-\Delta_\nu) \gamma(\tau)) R\biggl(\frac{2\pi
i|n-\Delta_\nu|}{c(c\tau +
  d)}\biggr) , \nonumber
\end{eqnarray}
Here $R(x)$ is the function
\be \label{eq:Rxint}
R(x):= 1-\frac{1}{\Gamma(1-w)}\int_x^\infty
e^{-z}z^{-w}dz = \frac{1}{\Gamma(1-w)}\int_0^x  e^{-z}z^{-w}dz \ee
The expression $F_\mu(\Delta_\mu)$ vanishes except when
$\Delta_\mu\in \mathbb{N}$, in which case it is  given by Eq.
(\ref{eq:constant}). We stress that
 Eq. (\ref{eq:sumcd}) is derived for general non-positive weight
  $w$, including integer and half-integer cases. The exclusion of
 positive weight is a consequence of the bound $p\geq 1$ in
 (\ref{eq:lipschitz}). Of course, the well-known technique of Poincar\'e
 series is applicable for $w>2$, since the sum is convergent in that
 case. Naive application of this technique for the reconstruction of a modular form with
 $w\leq 0$ from its polar part would not have the first term in (\ref{eq:sumcd}) and
would not have the regularizing factor $R(x)$. Note that the first
integral expression in (\ref{eq:Rxint}) shows that $R(x)$ approaches
$1$ exponentially fast for ${\rm Re}(x) \to \infty$, while the second
shows that $R(x) \sim \frac{x^{1-w}}{\Gamma(2-w)}$ for $x \to
0$. Using these simple estimates, convergence of the sum for $w\leq 0$
is established in Appendix \ref{sec:derivation}.

Eq. (\ref{eq:sumcd}) can be rewritten in the following form
\begin{eqnarray}
f_\mu(\tau)&=&\frac{1}{2}F_\mu(\Delta_\mu) + \frac{1}{2}\sum_{n-\Delta_\nu <0}
\lim_{K\to \infty} \sum_{(\Gamma_\infty
  \backslash \Gamma)_K} \\
&&M^{-1}(\gamma)_\mu^\nu F_\nu(n)
\left\{\frac{e((n-\Delta_\nu) \gamma(\tau))}{j(\gamma,\tau)^w} -
  r(\gamma,\tau,n-\Delta_\nu)\right\}. \nonumber
\end{eqnarray}

\noi For integer weight $r(\gamma,\tau,n-\Delta_\nu)$ can be simplified to
 \be
\label{eq:regintweight}
r(\gamma,\tau,n-\Delta_\nu)=\left\{\begin{array}{cc}\frac{e\left(
(n-\Delta_\nu) \frac{a}{c}\right)\sum_{l=0}^{|w|}\frac{1}{l!}\left(
\frac{2\pi i|n-\Delta_\nu|}{c(c\tau
    + d)}\right)^l  }{j(\gamma,\tau)^w}, & c\neq 0, \\
0, & c=0. \end{array}\right.
\ee
\noi This is the subtraction used in Refs. \cite{Denef:2007vg,
  Manschot:2007zb} to write a non-positive weight partition function
  directly as a fareytail. The same regularization had been previously
  used in the math literature in Ref. \cite{knopp:1990}. The
  generalization $R(x)$ is due to Niebur \cite{niebur1974}.

It is natural to ask if one can turn things around, that is:
starting with a projective representation $M(\gamma)$, and a
non-positive weight $w$, can one choose arbitrary coefficients
$F_\mu(n)$ with $n-\Delta_\mu<0$ and use Eq. (\ref{eq:sumcd}) to
construct a corresponding modular form with specified polar part? In
general, this is not possible.  We discuss this in   detail in
section \ref{sec:anomalies}, drawing on the technical results of
appendix \ref{sec:derivation}.

\subsection{Application to elliptic genera}
\label{subsec:ellgenera}
As explained in section \ref{sec:review}   elliptic genera may be
expressed as
  sums of theta functions with coefficients $h_\mu(\tau)$ forming a
  vector of modular forms.
The theta functions used in case of $\mathcal{N}=(2,2)$ elliptic
genera, transform as
\be \label{eq:thetatransform}
\theta_{m,\mu}(\tau,z)=M(\gamma)_\nu^\mu\frac{e\left(-m\frac{cz^2}{c\tau+
d}\right)}{j(\gamma,\tau)^{\frac{1}{2}}}\theta_{m,\nu}\left(\gamma(\tau),
\frac{z}{c\tau+d}\right). \ee

\noi We will insert Eqs. (\ref{eq:sumcd}) and
(\ref{eq:thetatransform}) in Eq. (\ref{eq:thetadecomposition}). The
coefficients $F_\mu(n)$ are in this case the Fourier coefficients of
the elliptic genus, $c(n,\ell) = c_\mu(4mn-\ell^2)$ with
$\ell=\mu\mod 2m$. Note that in this case $\Delta_\mu$ is given by
$\frac{\mu^2}{4m}\mod \mathbb{Z}$ and $F_\mu(\Delta_\mu)$ is only
non-zero when $\Delta_\mu\in \mathbb{N}$.

Thus we find for the elliptic genus of a Ricci flat manifold
\begin{eqnarray}
\label{eq:sumellipticgenus}
\chi(\tau,z)_X&=&\sum_{\mu \mod 2m}\frac{1}{2}
c_\mu (0)\theta_{m,\mu}(\tau,z) \\
&&+\frac{1}{2} \sum_{n-\frac{l^2}{4m}<0} \lim_{K\to \infty} \sum_{(\Gamma_\infty
  \backslash \Gamma)_K} c_\mu(4mn-l^2) \nonumber \\
&& \times e\left(n \gamma(\tau)+l\frac{z}{c\tau+d}-m\frac{cz^2}{c\tau +d }\right)R\left(\frac{2\pi i |n-\frac{l^2}{4m}|}{c(c\tau +
  d)}\right) \nonumber
\end{eqnarray}

\noi Note that we cannot write $c_\mu(0)\theta_{m,\mu}(\tau)$ as a
sum of simple exponential factors over $\Gamma_\infty \backslash
\Gamma$ but it could in principle
  be written as such a sum over $\Gamma_\infty \backslash \Gamma/ \Gamma_\infty$ by
Eq. (\ref{eq:constant}). Since the weight of the vector-valued modular
  forms is $-\frac{1}{2}$ in this case, $R(x)$ can be expressed as
\be
R(x)=\mathrm{erf}(\sqrt{x})-2\sqrt{\frac{x}{\pi}}e^{-x},
\ee
where $\mathrm{erf}(x)=\frac{2}{\sqrt{\pi}}\int_0^x e^{-t^2}dt$, which
is the error function.

Analogously, the $(0,4)$ elliptic genus can be written, using the
notation introduced below Eq. (\ref{eq:04eg}):
\begin{eqnarray}
\label{eq:04genus} \chi(\tau,   z)_{P}&=&\sum_{\mu   }
\frac{1}{2}H_\mu(\Delta_\mu)  \Theta_\mu(\tau,z)
+\frac{1}{2}\sum_{(\Gamma_\infty
  \backslash \Gamma)} \tilde
M^{-1}(\gamma) j(\gamma, \tau)^{\frac{3}{2}} j(\gamma, \bar
\tau)^{-\frac{1}{2}} E\left[-\frac{c}{c\tau+d}\frac{z^2}{2}\right]
\nonumber \\
&&\times \sum_{n,\mu:n-\Delta_\mu<0}H_\mu(n)
 R\left(\frac{2\pi i |n-\Delta_\mu|}{c(c\tau +
  d)}\right) e\left((n-\Delta_\mu)\gamma(\tau)\right)\nonumber \\
&&\times \sum_{q\in L_X + \mu^{\|} +P/2}  E\left[\half \gamma(\tau)
q^2 + q \cdot \left(\frac{z}{c\tau+d}+ \frac{P}{2}\right)\right].
\end{eqnarray}
 The exponentials of $\gamma(\tau)$ are weighted by the
quantity
\be n-\Delta_\mu - \half q_-^2. \ee
In the type IIA setting discussed in \cite{Denef:2007vg}  this
quantity  is the denoted $-\hat q_0$ and it can be written in terms
of D0- and D2-charges $(q_0, Q_a)$ using
\be \hat q_0 = q_0 - \half D^{ab} Q_a Q_b,\ee
where $D^{ab}$ is the matrix inverse of $D_{ab} = d_{abc} P^c$. In
this form, the polarity condition $\hat q_0>0$ is analogous to the
condition $n-l^2/4m<0$ in the $(2,2)$ case.

\section{Anomalies and Period Functions}
\label{sec:anomalies}\setcounter{equation}{0}

Let us now return to the question asked at the end of section
\ref{subsec:vector}. We have seen that the physical considerations
motivate the following problem in mathematics:

Suppose we are given a weight $w\leq 0$ and a rank $r$ multiplier
system $M(\gamma)$ on $\Gamma$. We wish to construct a vector-valued
modular form, transforming with weight $w$ and multiplier system $M$
with a prescribed polar part. That is,   the coefficients $F_\mu(m)$
in Eq. (\ref{fourexp}) with $m-\Delta_\mu<0$ are prescribed. Note that
consistency of this data requires $M(T^\ell)_\mu^\nu=e(-\delta_\mu
\ell) \delta_{\mu}^\nu$.

In general, there is an obstruction to finding such a vector-valued
form. We will show that the obstruction is measured by the
non-vanishing of a certain vector-valued cusp form of weight $2-w$ and multiplier
system $M(\gamma)^*$.

Let us begin by choosing a  vector $\delta$ with components
$\delta_\mu$, $\mu=1,\dots r$, some of whose components are
positive. We will attempt   to construct a vector-valued modular
form which behaves like
\be \label{eq:attempt} f(\tau) = \varepsilon(-\delta \tau)  +
 {\rm regular},
\ee
as $q\to 0$. Here $\varepsilon(-\delta \tau)$ is a vector with
components
\be \label{eq:defvareps} \varepsilon(-\delta \tau)_\mu  =
\begin{cases} e(-\delta_\mu \tau), & \delta_\mu >0, \\
0, & \delta_\mu \leq 0. \\ \end{cases} \ee
and ``regular'' means there is a $q$-expansion with non-negative
(possibly fractional) powers of $q$.

At first, it would appear to be straightforward to construct $f(\tau)$
by the method of images. Introduce the vector of functions
$$s_\gamma^{(\delta)}(\tau):= j(\gamma,\tau)^{-w} M(\gamma)^{-1} \varepsilon(-\delta
\gamma(\tau)).$$
Then it is elementary to check that
\be \label{eq:strmans} s_{\gamma\tilde \gamma}^{(\delta)}(\tau)
=j(\tilde\gamma,\tau)^{-w} M(\tilde\gamma)^{-1}
s_\gamma^{(\delta)}(\tilde \gamma \tau), \ee
and hence $s_{\gamma\tilde\gamma}^{(\delta)}(\tau) = s_{\tilde
\gamma}^{(\delta)}(\tau)$ for $\gamma \in \Gamma_\infty$.
Accordingly, we attempt to average:
\be \label{eq:naiveps} S^{(\delta)}(\tau) {\buildrel ? \over =}
\half \sum_{\gamma\in \Gamma_\infty\backslash \Gamma}
s_\gamma^{(\delta)}(\tau). \ee
Formally, from Eq. (\ref{eq:strmans}) we find
$S^{(\delta)}(\tilde\gamma \tau) = j(\tilde\gamma,\tau)^w M(\gamma)
S^{(\delta)}(\tau)$. Moreover, the   cosets $[\pm 1]$ lead to the
prescribed polar term and the remaining terms in the sum are regular
for $\tau \to i \infty$. It would thus appear that we have
succeeded, but in fact we have not.

The problem with the naive attempt Eq. (\ref{eq:naiveps}) is that for $c
\to \infty$ we have $|s_\gamma^{(\delta)}(\tau)| \sim |c\tau|^{- w}
$ and since we must have weight $w\leq 0$,  the series does not
converge. We therefore must regularize the series.

To motivate our regularization let us suppose  for the moment that
$-w\in \mathbb{N}$. We use the   identity
\be \label{eq:useful} \gamma(\tau) = \frac{a}{c} -
\frac{1}{c(c\tau+d)}, \ee
which is valid for $c\not=0$. This allows us to write
\be \label{eq:angs} e(-\delta\gamma(\tau)) = e^{-2\pi i \delta
\frac{a}{c}} e^{2\pi i \frac{\delta}{c (c\tau+d)}}. \ee
An evident regularization would be to subtract the first $\vert w
\vert$ terms from the Taylor series expansion of $e^{2\pi i
\frac{\delta}{c (c\tau+d)}}$ around zero.  Thus we introduce the
regularized sum:
\be \label{eq:Regps} S^{(\delta)}_{\rm Reg}(\tau) :=\half
\sum_{\gamma\in \Gamma_\infty\backslash \Gamma}
(s_\gamma^{(\delta)}(\tau)+ t_\gamma^{(\delta)}(\tau)), \ee
with
\be \label{eq:tdef} t_\gamma^{(\delta)}(\tau):= -j(\gamma,\tau)^{-w}
M^{-1}(\gamma)  \sum_{j=0}^{\vert w\vert} {1\over j!} \biggl(
\frac{1}{c (c\tau+d)}\biggr)^j \left(2\pi i \delta\right)^j e^{-2\pi
i \delta \frac{a}{c}} . \ee
Here and in what follows we understand expressions like $\left(2\pi
i \delta\right)^j e^{-2\pi i \delta \frac{a}{c}}$ to be vectors
whose $\mu^{th}$ component is zero if $\delta_\mu\leq 0$ and is
$\left(2\pi i \delta_\mu\right)^j e^{-2\pi i \delta_\mu
\frac{a}{c}}$ if $\delta_\mu>0$, as in  Eq. (\ref{eq:defvareps}).
Note that $t_\gamma^{(\delta)}(\tau)$ is a polynomial in $\tau$.
Moreover, the sum in Eq. (\ref{eq:Regps}) is
convergent.\footnote{The convergence is actually a little delicate.
One must group together terms with positive and negative values of
$d$ to avoid a logarithmic divergence in the sum over $d$. Once this
is done, convergence can be shown for $w \leq 0$. See Appendix A for
more details. }

Now, the regularization has been carried out for $w$ integral.
Remarkably, it may be generalized to non-integral $w$ as follows.
Returning to the expression for $ t_\gamma^{(\delta)}(\tau)$ we
recognize a truncated exponential series. The latter can be written
in terms of the incomplete Gamma function using the identity (see
Eq. (\ref{hfunction}) below):
\be \sum_{k=0}^\infty \frac{x^{k+1-w}}{\Gamma(k+2-w)} = e^x \biggl(
1- \frac{1}{\Gamma(1-w)} \int_x^{\infty} e^{-z} z^{-w} dz \biggr).
\ee
Using this we may write $t_\gamma^{(\delta)}(\tau)=0$  for $c=0$,
while for $c\not=0$,
\be \label{Regtrm} t_\gamma^{(\delta)}(\tau) := -
j(\gamma,\tau)^{-w} M^{-1}(\gamma) \varepsilon(-\delta\gamma(\tau))
 \frac{1}{\Gamma(1-w)}
\int_{x(\gamma,\delta)}^\infty e^{-z}z^{-w}dz,\ee
where the factor multiplying $M^{-1}(\gamma)$ on the right is the
vector whose $\mu^{th}$ component is zero for $\delta_\mu\leq 0$ and
\be e(-\delta_\mu \gamma(\tau))
 \frac{1}{\Gamma(1-w)}
\int_\frac{2\pi i \delta_\mu}{c(c\tau +d)}^\infty e^{-z}z^{-w}dz,
\ee
for $\delta_\mu>0$.   In this form the regularization Eq. (\ref{Regtrm})
still makes sense for $w$ non-integral, and the regularized sum is
again convergent. This follows from the $x\to 0$ asymptotics of
$R(x)$.

Of course, now our regularization has spoiled the formal covariance
under modular transformations! However, it turns out that it has
spoiled it in a controlled way because $t_\gamma^{(\delta)}(\tau)$
is related to certain \textit{period integrals}.  For any function
$h(\tau)$ on $\CH$ decaying sufficiently rapidly at ${\rm Im}(\tau)
\to \infty$ we can define its period function
\be p(\tau, \bar y, \bar h) := \frac{1}{\Gamma(1-w)} \int_{\bar
y}^{-i\infty} \overline{h(z)} (\bar z -\tau)^{-w} d \bar z. \ee
Then we claim that
\be t^{(\delta)}_\gamma(\tau) = p(\tau, \gamma^{-1}(-i\infty),
\overline{g^{(\delta)}_\gamma}), \ee
where
\be g^{(\delta)}_\gamma(z) := j(\gamma,z)^{w-2}
\overline{M^{-1}(\gamma)} (-2\pi i \delta)^{1-w} \varepsilon(\delta
\gamma(z)). \ee
Now, $g^{(\delta)}_\gamma(z)$ transforms simply, and from this one
can verify that
\be \label{eq:teeshift} t^{(\delta)}_\gamma(\tilde\gamma \tau)=
j(\tilde \gamma,\tau)^w M(\tilde \gamma) \biggl[
t^{(\delta)}_{\gamma\tilde\gamma}(\tau)- p(\tau,
\tilde\gamma^{-1}(-i\infty),\overline{g^{(\delta)}_{\gamma\tilde\gamma}})\biggr].
\ee
Because of the second term in Eq. (\ref{eq:teeshift})  our regularized
sum does not transform covariantly. Rather we have:
\be \label{eq:regStransform} S^{(\delta)}_{\rm Reg}(\tilde
\gamma\tau) = j(\tilde \gamma,\tau)^w M(\tilde
\gamma)S^{(\delta)}_{\rm Reg}(\tau) -
 j(\tilde \gamma,\tau)^w
M(\tilde \gamma)\half \sum_{\Gamma_\infty\backslash\Gamma} p(\tau,
\tilde\gamma^{-1}(-i\infty),\overline{g^{(\delta)}_{\gamma\tilde\gamma}}).
\ee
Now we would like to simplify the ``anomalous'' second term on the
right-hand side of Eq. (\ref{eq:regStransform}). To this end we would
like to exchange the summation with the integration in the
definition of the period function. Although the second term involves
an absolutely convergent sum, we must be very careful about
exchanging the sum and integration as well as redefining the sum by
$\gamma \to \gamma\tilde\gamma^{-1}$. Using results of Niebur
\cite{niebur1974}, which are further explained  in the appendix, we
have:
\be \half \sum_{\Gamma_\infty\backslash\Gamma} p\left(\tau,
\tilde\gamma^{-1}(-i\infty),\overline{g^{(\delta)}_{\gamma\tilde\gamma}}\right)=
p\left(\tau, \tilde\gamma^{-1}(-i \infty),
\overline{G^{\delta}}\right) + j(\tilde \gamma, \tau)^{-w} M(\tilde
\gamma)^{-1} F(\delta) - F(\delta), \ee
where
\be G^{(\delta)}(\tau):=\half \sum_{\Gamma_\infty\backslash \Gamma}
g^{(\delta)}_{\gamma}(\tau), \ee
and $F(\delta)$ is a vector of constants given by
\be
 F(\delta)_\mu = \begin{cases}  \pi \sum_{\delta_\nu>0}\frac{(2\pi
  \delta_\nu)^{1-w}}{\Gamma(2-w)} \sum_{c=1}^\infty
  c^{w-2}K_c(0_\mu,-\delta_\nu), & \delta_\mu \in \mathbb{N},\\ 0, &
  \delta_\mu \not\in \mathbb{N}. \\ \end{cases}
\ee
where $0_\mu$ is the vector all of whose components are zero and
$K_c$ is the generalized Kloosterman sum of Eq. (\ref{eq:kloosterman}).

 The net
result of all of this is that in our attempt to construct the weight
$w$ modular vector with polar term (\ref{eq:attempt}) the method of
images leads us - more or less uniquely -- to define a vector of
functions $\hat S^{(\delta)}_{\rm Reg}(\tau):= F(\delta) +
S^{(\delta)}_{\rm Reg}(\tau)$. As $\tau \to i \infty$ this vector
indeed behaves as
\be \hat S^{(\delta)}_{\rm Reg}(\tau) =  \varepsilon(-\delta \tau) +
{\rm regular}. \ee
However, it satisfies the  transformation law:
\be \label{eq:truetransform} \hat S^{(\delta)}_{\rm Reg}(\tilde
\gamma \tau)= j(\tilde \gamma,\tau)^w M(\tilde \gamma) \biggl[\hat
S^{(\delta)}_{\rm Reg}(\tau) - p^{(\delta)}(\tau,\tilde \gamma)
  \biggr],
\ee
where
\be \label{eq:periodint} p^{(\delta)}(\tau,\tilde\gamma):=
p(\tau,\tilde\gamma^{-1}(\infty),\overline{G^{(\delta)}}) =\frac{1}{\Gamma(1-w)}
\int_{-\tilde d/\tilde c}^{-i \infty} \overline{G^{(\delta)}(z)}
(\bar z - \tau)^{-w} d \bar z, \ee
is a vector of functions defined by
\be \label{cuspgee} G^{(\delta)}_{\mu}(z) = \half\sum_{\gamma\in
\Gamma_\infty\backslash \Gamma} j(\gamma,z
)^{w-2}\sum_{\delta_\nu>0} \left(M^{-1}(\gamma)_\mu^\nu\right)^*
(-2\pi i \delta_\nu)^{1-w} e(\delta_\nu \gamma(z)). \ee
The vector of functions $ p^{(\delta)}$ is an obstruction to the
existence of $f(\tau)$.

In contrast to Eq. (\ref{eq:naiveps}), the series (\ref{cuspgee})
for $G^{(\delta)}(z)$ is nicely convergent. It therefore follows
that $G^{(\delta)}(\tau)$ is a vector-valued modular form of weight
$2-w$ transforming according to
\be G^{(\delta)}(\gamma \tau)  =
j(\gamma,\tau)^{2-w}\overline{M(\gamma)}G^{(\delta)}( \tau). \ee
In fact, $G^{(\delta)}$ is a vector-valued \textit{cusp form}, that
is, the components vanish for $\tau \to i \infty \cup
\Gamma(i\infty)$. This follows since it is clear from the series
expansion that $G^{(\delta)}$ vanishes for $\tau \to i \infty$.  We
give an explicit formula for the Fourier coefficients in Eq.
(\ref{fouriercoeffs}) below.

Lemma 3.2 of Ref. \cite{niebur1974} shows that the period integral
vanishes if and only if $G^{(\delta)}$ vanishes. Therefore,  our
cusp form  $G^{(\delta)}$, if non-vanishing, forms  an obstruction
to constructing the vector valued form with prescribed polar term
$\delta$.  The shift in Eq. (\ref{eq:truetransform}) by
$p^{(\delta)}$ represents an anomaly under modular transformations.
This is a familiar situation in quantum field theory:  a divergent
quantity is formally invariant, the regularized quantity breaks the
invariance, but in a controlled way. Thus the problem of
constructing a true modular form with negative weight and specified
polar part is a kind of anomaly cancellation problem: one must form
linear combinations $\sum_\delta \Omega_\delta \hat
S^{(\delta)}_{\rm Reg}$ so that the associated cusp form cancels.
The coefficients $\Omega_\delta$ are exactly the ``polar
degeneracies'' that play a crucial role in the physical discussions
of the fareytail transform and the OSV conjecture.

In fact, the analogy goes deeper, since the anomaly is in fact
related to a cohomology theory known as Eichler cohomology. It
follows from the definition of the period vector that we have the
transformation law given in Eq. (\ref{eq:transformp}). Therefore
\be p^{(\delta)}( \tau, \tilde \gamma)- p^{(\delta)}(\tau, \gamma
\tilde \gamma )+ j(\tilde\gamma,\tau)^{-w} M(\tilde\gamma)^{-1}
p^{(\delta)}( \tilde \gamma\tau,\gamma)=0. \ee
Defining the standard slash operator on functions $f(\tau,\gamma)$:
\be f(\cdot,\gamma)\vert^M_{w}\tilde\gamma :=
 j(\tilde\gamma,\tau)^{-w}
 M(\tilde\gamma)^{-1}f(\tilde\gamma\tau,\gamma),
 \ee
we see that the obstruction to modularity lies in the space of
functions satisfying
\be \label{cocycle} f(\cdot, \tilde \gamma) - f(\cdot,
\gamma\tilde\gamma) + f(\cdot,\gamma)\vert_w^M\tilde\gamma =0.  \ee
If we interpret $f(\tau,\gamma)$ as a cochain on the group $\Gamma$
with values in functions of $\tau$ then (\ref{cocycle}) is the
statement that $f$ is a 1-cocycle. A $1$-coboundary is a function of
the form  $f(\cdot,\gamma) = b(\cdot) - b(\cdot)\vert_w^M\gamma$
where $b(\tau)$ is a single function of $\tau$.  We would like to
define a cohomology group as $1$-cocycles modulo $1$-coboundaries.
Of course, the transformation law  (\ref{eq:truetransform}) shows
that $\hat S^{(\delta)}_{\rm Reg}$ trivializes $p^{(\delta)}$, so to
get an interesting theory we need to restrict the $\Gamma$ module of
functions in which we compute cohomology.

When the weight $w$ is a negative integer,
$p^{(\delta)}(\tau,\gamma)$ is a vector of polynomials of degree
$\leq \vert w\vert$. In the scalar case the space of obstructions to
constructing a modular form with prescribed polar part is
$H^1(\Gamma, V_{\vert w \vert})$ where $V_{\vert w\vert}$ is the
vector space of polynomials of degree $\leq \vert w\vert$. For $
\vert w\vert \not\in \mathbb{N}$, we are forced to work in a larger
space of functions, those with at most polynomial growth at the
cusps.   We refer to references \cite{knopp:1990, knopp:1974,
zagier:1991} for more details.

We conclude by giving some more explicit conditions on the polar
degeneracies $\Omega_\delta$ for anomaly cancellation. Note first
that $p^{(\delta)}(\tau,T)=0$ so it suffices to check
\be \sum_\delta \Omega_\delta p^{(\delta)}(\tau,S)=0 \ee
since $S,T$ generate $\Gamma$. In the case of $-w \in \IN$ the
coefficients of such a period polynomial are calculated by
$\int_0^{i\infty} G^{(\delta)}(z)z^{s-1}dz$, $s\in \mathbb{N}$. Such
integrals are known as Mellin transforms. When the Fourier expansion
of $G^{(\delta)}(\tau)$ is given by
\be
G^{(\delta)}(\tau)_\mu=\sum_{n+\alpha_\mu>0}u^{(\delta)}(n)_{\mu}
q^{n+\alpha_\mu}, \qquad \alpha_\mu=\delta_\mu-\lfloor \delta_\mu
\rfloor, \ee
then the Mellin transform $\mathcal{M}(G^{(\delta)},s)$  can be
calculated to be
\be \mathcal{M}(G^{(\delta)},s)=\frac{\Gamma(s-1)}{(-2\pi i)^s}
\sum_{n+\alpha>0}^\infty\frac{u^{(\delta)}(n)}{(n+\alpha)^s}. \ee
These quantities can be analytically continued to general values of
$s$. Series like $\sum_{n+\alpha>0}^\infty\frac{a(n)}{(n+\alpha)^s}$
are known as {\it L}-series. Thus, anomaly cancellation can be
expressed in terms of $L$-series.

In the case of $w$ half-integral the period functions are much more
complicated than polynomial, but can be expressed in terms of error
functions. For example, for $w=-1/2$
\be \overline{p^{(\delta)}(\tau)_\mu } = \frac{e^{3\pi
i/4}}{\Gamma(3/2)} \sum_{n+\alpha_\mu>0}
\frac{u^{(\delta)}_\mu(n)}{(2\pi (n+ \alpha_\mu))^{3/2}}
e((n+\alpha_\mu)\bar \tau) \Gamma\left(\frac{3}{2}, 2\pi i (n+\alpha_\mu)
\bar\tau \right).\ee
The upper incomplete Gamma function can be written as
\be \Gamma(3/2,x) = x^{1/2} e^{-x} + \frac{\sqrt{\pi}}{2}{\rm
erfc}(\sqrt{x}), \ee
where $\mathrm{erfc}$ is the complementary error function,
$\mathrm{erfc}(x)=1-\mathrm{erf}(x)$.

Returning to the case of general weight, for   completeness we give
the Fourier decomposition of $G^{(\delta)}$:
\begin{eqnarray}\label{fouriercoeffs}
G^{(\delta)}_\mu(\tau) & = &   (-2\pi i
\delta_\mu)^{1-w}e(\delta_\mu
\tau) \theta(\delta_\mu>0)  \\
&& + i (-2\pi i)^{2-w}  \sum_{\ell+\delta_\mu>0}
e((\ell+\delta_\mu)\tau) \Biggl\{
\sum_{c=1}^\infty\sum_{\delta_\nu>0} \frac{1}{c}\tilde
K_c(\ell+\delta_\mu,\delta_\nu) \nonumber \\
&&\times   \left(\delta_\nu (\ell+\delta_\mu)\right)^{(1-w)/2}
J_{1-w}\left(\frac{4\pi}{c}\sqrt{(\ell+\delta_\mu)\delta_\nu}
\right)\Biggr\}, \nonumber
\end{eqnarray}
with generalized Kloosterman sum
\be \tilde K_c(\ell+\delta_\mu,\delta_\nu)= e^{-i \pi (2-w)/2}
\sum_{0\leq d < c; (d,c)=1} e((\ell+\delta_\mu)\frac{d}{c})
\left(M^{-1}(\gamma_{c,d})_\mu^\nu\right)^*
e(\delta_\nu\frac{a}{c}). \ee
This is a straightforward application of the Poisson summation
formula.

Besides calculation of the Fourier coefficients of
$G^{(\delta)}(\tau)$ directly, a decomposition of
$G^{(\delta)}(\tau)$ in terms of a basis of cusp forms is
instructive as well. This is potentially useful since we have
learned that the obstruction to forming a good modular form with
prescribed polar term lies in a space isomorphic to the space of vector-valued
cusp forms $S\left(2-w, \overline{M}\right)$.   Let us restrict
attention to the scalar case for simplicity. We denote an orthonormal basis of
the appropriate cusp forms by $H^j(\tau)$, with $j=1 \dots
\dim\left[ S\left(2-w, \overline{M}\right)\right]$. The Fourier
coefficients of $H^j(\tau)$ are defined  by
$$ H^j(\tau) = \sum_{n\geq 0} h^j(n) q^{n+\alpha}$$
with $\alpha = \delta - \lfloor \delta \rfloor$.   The Petersson
inner product calculates the coefficients of $G^{(\delta)}(\tau)$
with respect to this basis. By unfolding of the integration domain
we find
\begin{eqnarray}
\int_{\Gamma\backslash
  \mathcal{H}}G^{(\delta)}(\tau)\overline{H^j(\tau)}y^{2-w}\frac{dx
  dy}{y^2}=\frac{\Gamma(1-w)}{(2i)^{1-w}}\overline{h^j(\lfloor \delta \rfloor)},
\end{eqnarray}
where $x$ and $y$ are respectively the real and imaginary part of
$\tau$. The question whether a given set of polar terms  gives rise
to a vector-valued modular form is now reduced to the finite set of
conditions:
\be \label{eq:cuspanomcan} \forall j\qquad  \sum_{ \delta>0 }
\Omega_\delta \overline{h^j(\lfloor \delta \rfloor)}=0 . \ee
This is a difficult question to analyze in general, but is
potentially tractable for the cases when a concrete basis of
$S\left(2-w, \overline{M}\right)$ is known. \footnote{As a measure
of the difficulty involved suppose the weight $w=-10$. In this case
$h^i(\lfloor \delta \rfloor) = \tau(\delta)$ are the famous
Ramanujan functions. We are trying to construct integral linear
combinations of these coefficients which vanish.} 

In the case of $(2,2)$ elliptic genera, we have to consider
vector-valued cusp forms. These vector-valued cusp forms can be mapped
to scalar cusp forms of congruence subgroups \cite{eichler} with
weight $2-w$. The dimension of the spaces of these cusp forms is
expected to grow linearly in $m$ \cite{Cohen:1977}. A more precise
study shows that the space of obstructions can be related to a proper
subspace of the space of cusp forms known as the Kohnen $+$-space
\cite{Kohnen:1982}.

In the case of $(0,4)$ elliptic genera as we scale $P \to \lambda P$ a
rough estimate suggests the number of polar terms scales as
$\lambda^{b_2+3}$, whereas the dimension of the space
of relevant cusp forms scales only as $\lambda^{b_2}$. We refer to
Ref. \cite{Manschot:2008zb}, were  a more precise calculation of these
quantities is performed.

\section{Applications of the Fareytail Expansion}
\label{sec:applications}\setcounter{equation}{0}

\subsection{The fareytail transform revisited}
\label{subsec:fareyrevisited}
We now put into the present perspective the discussions of the
fareytail transform which have appeared previously in
Ref. \cite{Dijkgraaf:2000fq,Moore:2004fg}.

First, the transformation law (\ref{eq:modtransform})  makes clear
why the fareytail transformation is flawed in general. In the
present context we would use the operator $\CO =
\bigl(q\frac{d}{dq}\bigr)^{1-w}$ which formally transforms modular
forms of weight $w$ to modular forms of weight $2-w$. Being a
(pseudo-)differential operator it cannot change the multiplier
system $M(\gamma)$. On the other hand, substituting $\gamma=-1$ in
Eq. (\ref{eq:modtransform}) we find $ f_\mu(\tau) = e^{-i \pi w}
M(-1)_\mu^\nu f_\nu(\tau)$. Since  the $f_\mu(\tau)$ are independent
functions of $\tau$    we conclude that $ M(-1)_\mu^\nu= e^{i \pi w}
\delta_\mu^\nu$. Since the multiplier system does not change under
the fareytail transform we must have $ e^{i \pi w } = e^{i \pi
(2-w)}$ implying $e^{2\pi i w } =1 $ implying that $w$ is integral.
\footnote{The reason adduced by Don Zagier for the failure of the
transform for $w$ half-integral was based on results concerning the
field of definition of the Fourier coefficients of modular forms.}

On the other hand, the fareytail transform is valid in the case of
non-positive integer weight. We summarize the arguments from Ref.
\cite{Dijkgraaf:2000fq,Moore:2004fg}. In this  case the operator
$\left(q\frac{d}{dq}\right)^{1-w}$ really does map a modular form of
weight $w$ to a   modular form of weight $2-w$ thanks to Bol's
identity
\begin{equation} \label{coolidentity}
 L^n \left[ (c \tau + d)^{-1+n} f\left( \frac{a \tau + b}{c \tau +
d} \right)
 \right] = (c\tau+d)^{-1-n} (L^n f)\left( \frac{a \tau + b}{c \tau + d}
 \right),
\end{equation}
where $L:=q{d\over dq}$. Bol's identity is valid for any non-negative
integer $n$ and any suitably differentiable function $f(\tau)$.
If $f(\tau)$ is a modular form of weight $w$ and with a pole for
$q\to 0$, we define $\tilde f(\tau) := \CO f(\tau)$.  Using a
regularized Petersson inner product one shows that $\tilde f(\tau)$
is orthogonal to nonsingular modular forms and is hence uniquely
determined by its polar part \cite{Moore:2004fg}. Therefore, the convergent Poincar\'e
series of weight $2-w$ obtained by averaging the polar part of
$\tilde f$ must in fact be equal to $\tilde f$.

In terms of the considerations of section \ref{sec:anomalies} the
orthogonality to nonsingular modular forms is now interpreted as our
anomaly cancellation condition on the polar part of $f$. Moreover,
taking the case of trivial multiplier system for simplicity,  we
have
\begin{eqnarray}
 \CO \sum_{\Gamma_\infty\backslash \Gamma} \sum_{\delta>0} \Omega_\delta
(s_\gamma^{(\delta)} + t_\gamma^{(\delta)} ) & =&
\sum_{\Gamma_\infty\backslash \Gamma} \sum_{\delta>0} \Omega_\delta
\CO\biggl[ (c\tau+d)^{-w} \exp\left(-2\pi i \delta
\frac{a\tau+b}{c\tau+d}\right)\biggr]\\
&  = & \sum_{\Gamma_\infty\backslash \Gamma} \sum_{\delta>0}
\Omega_\delta \left.\biggl( \CO e(-\delta \tau)
\biggr)\right|_{2-w}\gamma, \nonumber
\end{eqnarray}
where in the first line we can exchange summation and
differentiation on the left-hand side, but not on the right-hand
side.  The operator $\CO$ annihilates the constant term in the
Poincar\'e series as well as the regularizing term
$t^{(\delta)}_\gamma$ (since the latter is a polynomial in $\tau$ of
order $\vert w \vert$). We have used Bol's identity to write the
second line. The second line is indeed the claimed Poincar\'e series
expansion of the polar part of $\tilde f$. Thus, we have recovered
the previous story.

\subsection{AdS/CFT interpretation}
The introduction motivated the Poincar\'e series as a sum over
classical geometries. We have seen that this semi-classical expansion is
remarkably accurate for the partition functions of BPS states. The
sums given by Eqs. (\ref{eq:sumcd}), (\ref{eq:sumellipticgenus}) and
(\ref{eq:04genus}) are however more involved than the gravity path integral
described in the introduction. The elliptic genera contain a theta function and the polar part can
possibly consist of many terms. We will briefly discuss these
aspects here and point out a subtlety with respect to the constant
term of the partition function. This subtlety is new since the
fareytail transform, present in previous discussions, would
annihilate the constant term.

The dependence on $z$ in Eq. (\ref{eq:sumellipticgenus}) is a
consequence of the fact that we are not dealing with pure gravity but
with a reduction of Type IIB string theory to AdS$_3$. The parameter $z$
 arises since the bulk contains $SU(2)$ gauge fields. It corresponds to a Wilson
line from the three-dimensional point of view
\cite{Dijkgraaf:2000fq}. States in the bulk are also well described
in six-dimensional supergravity on AdS$_3\otimes S^3$.  The $z$
variable couples then to the momentum of spinning particles on the
$S^3$. In the $(0,4)$ elliptic genus the  parameters $y$ arise
similarly from the presence of a number of $U(1)$ gauge fields in
the bulk.

Eq. (\ref{eq:sumellipticgenus}) contains a sum over
$n-\frac{l^2}{4m}<0$. The contribution of these states in thermal
AdS$_3$ to the full elliptic genus, is given by
\be \chi(\tau,z)^-=\sum_{{-m+1\leq \mu \leq m  \atop
    4mn-\mu^2<0}}c_\mu(4mn-\mu^2) q^{n-\frac{\mu^2}{4m}}
\theta_{m,\mu}(\tau,z).
\ee
\noi This partition function counts only the ``light'' excitations of thermal
AdS$_3$. These excitations are typically Kaluza-Klein modes or
(charged) point particles. The charged point particles can be branes
wrapping cycles in an orthogonal compact manifold. The theta function
arises from the singleton modes. The cut-off on the
contributing states appears to be equal to the cosmic censorship bound
for black holes. This bound is given by $4mM-J_0^2\geq 0$ with
$M=L_0-\frac{c_L}{24}$ \cite{Cvetic:1998xh}. The ``light'' excitations
are thus exactly those states which do not collapse to a black
hole in thermal AdS$_3$. This is the regime where counting of the
degeneracies in supergravity could be reliable. For a meaningful
comparison between supergravity and CFT, we apply spectral flow to
transform the trace over the R-R sector to the NS-NS sector. To avoid
confusion we will denote the eigenvalues of $L_0-\frac{c_L}{24}$ in
the NS sector by $n_\mathrm{NS}$. Refs. \cite{deBoer:1998us,
  Maldacena:1999bp} have shown that the supergravity degeneracies
indeed match with the CFT degeneracies for small values of
$n_\mathrm{NS}$, in particular $n_\mathrm{NS}< 0$ . The computations
on either side of the correspondence do not match for states with a
higher energy. This suggests that gravitational degrees of freedom
start contributing at this level. Since $n_{\mathrm{NS}}=0$ is the
smallest value of $n_{\mathrm{NS}}$ which satisfies the cosmic
censorship bound this is not surprising \cite{Dijkgraaf:2000fq}. The
fareytail expansion of the elliptic genus (Eq.
(\ref{eq:sumellipticgenus})) is a sum of the light excitations in
all the black hole geometries. The excitations which would collapse
into the black hole are excluded, since those states are counted by
another classical black hole geometry in the sum.

The exponent of the classical action is multiplied by $R\left(\frac{2\pi i
  |n-\frac{l^2}{4m}|}{c(c\tau+d)}\right)$. As explained in depth in
previous sections, this factor is indispensable for a proper convergence of the
gravity path integral. Moreover it has the effect of a smooth cut-off on the
contributions of the light excitations in thermal AdS$_3$ to the
geometries with $c\neq 0$, since $R\left(\frac{2\pi i
  |n-\frac{l^2}{4m}|}{c(c\tau+d)}\right)$ is exponentially close to 1
for $|n-\frac{l^2}{4m}|\gg 1$, and is zero for $|n-\frac{l^2}{4m}|=0$. The geometries with complicated
topologies ($c$ and/or $d\gg 1$) are similarly cut-off.

We would like to draw attention now to the contribution to the
elliptic genus of states with $4mn-l^2=0$. Half of these states are
counted by the term
$$\sum_{\mu \mod
  2m}\frac{1}{2}c_\mu(0)\theta_{m,\mu}(\tau, z),$$
   which   appears separately in
Eq. (\ref{eq:sumellipticgenus}). Comparison with the Fourier series
of the elliptic genus, Eqs. (\ref{eq:fourierjacobi}) and
(\ref{eq:fourierjacobidecomp}), shows that the sum over
$\Gamma_\infty\backslash \Gamma$ contains an equal term. This
suggests that half of the states at $n-\frac{l^2}{4m}=0$ correspond
to black holes, whereas the other half are stable states in thermal
AdS$_3$. Since these stable states in thermal AdS$_3$ do not
contribute to the black hole states, their interpretation is more
subtle than the states with $4mn-l^2<0$. The way the states at the
threshold appear in the partition function leads us to suggest that
these excitations are so close to a collapse
 in thermal AdS$_3$, that they would collapse into the
black hole when added to a black hole geometry. A more quantitative
description of this phenomenon is highly desirable.

At a heuristic level the factor   $R\left(\frac{2\pi
i|n-\frac{l^2}{4m}|}{c(c\tau+d)}\right)$ can be understood in a
similar way as the ``fraction'' of light excitations with a given
value of $4mn - l^2$ in thermal AdS$_3$, which can exist as a stable
excitation of the black hole given by $(c,d)$. The other states are
unstable and will collapse into the black hole. Note that this
quantity is in general complex so such an interpretation is
heuristic, at best.

The polar states in the case of the $\mathcal{N}=(0,4)$ elliptic
genus have a similar interpretation of states which are not massive
enough to form black holes. They include massless supergravity modes
as well as M2-branes and anti-M2-branes \cite{Gaiotto:2006ns}. In
addition there are other exotica such as M5-black rings, $\IZ_r$
quotients of $AdS_3 \times S^2$ and even more complicated
geometries. We expect these are all dual to the multi-centered
D6 anti-D6 configurations that played a crucial role in Ref.
\cite{Denef:2007vg}.

Finally, we comment on an ambiguity related to the Poincar\'e
series. We have argued that the states counted by the theta function
are pure gauge in the bulk and only dynamical on the
boundary. Therefore, these states should not be summed over all
different bulk geometries. This interpretation implies that all
non-polar states are black hole states. This statement might be
questioned for the following reason. The singleton degrees of freedom
are not just given by the theta function, since these enumerate only
the primaries. The descendants of the primaries should also be
included, since they are also excitations on the boundary, and not to
be summed over all geometries. In addition, Ref. \cite{Witten:2007kt}
explains that the descendants of primaries should not be considered as
black hole states. Since the descendants are
not black hole states, one should sum these descendants over all
geometries. In other words, in the Poincar\'e series for $f_\mu(\tau)$
 one wants to remove the condition $n-\Delta_\mu<0$  and include also
the descendants of the polar primaries.  

Except for a special situation, this does not seem to be allowed by the
analysis of this paper, since the non-polar terms lead to non-vanishing obstruction
forms with a polar part. However in the case of weight 0, and trivial
multiplier system, meromorphic obstruction forms can be written as the derivative of a meromorphic
weight zero form, such that the integrand of the period function is a
total derivative. Since the boundary of the integration domain are two
equivalent cusps under $\Gamma$, the modular anomaly vanishes. Also
non-polar terms can therefore be included in the Poincar\'e series
without affecting modularity. Unfortunately, we are not aware of a
generalization to the vector-valued case.

%
%

\subsection{Phase transitions}
One attractive feature of the fareytail expansion is that it is
well-suited to deduce phase transitions between different AdS$_3$
geometries \cite{Dijkgraaf:2000fq}. Such phase transitions were
first described in four dimensions by Hawking and Page
\cite{Hawking:1982dh} and interpreted in the AdS/CFT context by
Witten \cite{Witten:1998zw}. We can understand the phase
transformations by determining which term in the sum
(\ref{eq:gravpathintegral}) contributes most to the partition
function. We have \be
|Z_\mathrm{grav}(\tau)| \leq \sum_{\Gamma_\infty \backslash \Gamma}
e^{\frac{2\pi c_L}{24}\frac{\im(\tau)}{|c\tau+d|^2}}. \ee

\noi So the combination of $(c,d)$ which maximizes
$\frac{\im(\tau)}{|c\tau +d|^2}$ determines the term which
  contributes most to the path integral. This $(c,d)$ describes the
  dominant   classical geometry. Phase transitions occur between
  geometries by variation of $\tau$. The regularizing factor $R\left(\frac{2\pi i
    |n-\Delta_\nu|}{c(c\tau+d)}\right)$ does not change this
  conclusion. To see this we estimate $\left|R\left(\frac{2\pi i
    |n-\Delta_\nu|}{c(c\tau+d)}\right)-1\right|$:
\be \label{eq:rbeta} \left|R\left(\frac{2\pi i
  |n-\Delta_\nu|}{c(c\tau+d)}\right)-1\right| \leq \frac{1}{\Gamma (1-w)}
\left(\frac{2\pi
  \frac{c_L}{24}}{|c(c\tau+d)|}\right)^{w-1}e^{-2\pi\frac{c_L}{24}\frac{\im(\tau)}{|c\tau+d|^2}},
\ee

\noi where we assumed that $\frac{2\pi |n-\Delta_\nu|}{|c(c\tau+d)|} \gg 1$. We observe that the
correction is typically exponentially smaller than the exponent of
the classical action, and we can conclude that the new fareytail
predicts as well phase transitions parametrized by
$\Gamma_\infty\backslash \Gamma$.

\subsection{The OSV conjecture}
\label{subsec:OSVconjecture}

The  fareytail expansion of $(0,4)$ elliptic genera has been used in
recent attempts to prove a refined version of the OSV conjecture
\cite{ooguri:2004zv, deBoer:2006vg,Denef:2007vg}. The regularization
factor $R(x)$ does not alter the discussion when the black hole
charges are such that the saddle point topological string coupling
is strong. In the notation of Ref. \cite{Denef:2007vg} we have
\be g_s \sim \sqrt{\frac{-\hat q_0 }{P^3}} \gg 1. \ee
The dominant term in the evaluation of $\Omega(\mathcal{Q})$, where
$\mathcal{Q} = P + Q + q_0 dV$ is the charge of a D4-D2-D0 brane
system on a Calabi-Yau manifold $X$, is the $c=\pm 1,d=0$ term in
the fareytail expansion of the $(0,4)$ elliptic genus for $\tau
\cong i \sqrt{P^3/\vert \hat q_0\vert}$. Therefore, for strong
topological string coupling   ${\rm Re}(x)\to \infty$ in the
argument of $R(x)$. Thus the regularization factor introduces
exponentially small corrections in this regime. In this way the
artificial restriction to $b_2(X)$ even, imposed in Ref.
\cite{Denef:2007vg}, may be removed.

On the other hand, in the more interesting regime of \textit{weak}
topological string coupling, $P^3\gg \vert \hat q_0\vert$ the value
of $x$ goes to zero for the $c=\pm 1,d=0$ terms in the fareytail
expansion and the effects of our regularization become significant,
introducing further corrections to the OSV formula in this regime.

An interesting phenomenon described  in Ref.
\cite{Denef:2007vg,Denef:2007yn} is the ``entropy enigma.'' This
refers to the fact that for charges corresponding to weak
topological string coupling, semi-classical multicentered
states exist which contribute to the ``large radius BPS degeneracies''
$\Omega(\CQ)$ with entropies which grow exponentially in $P^3$ for
$P\to \infty$. In particular, they dominate the single centered
entropy, the latter growing   like $\sqrt{-\hat q_0 P^3}$. A growth
of $\log\vert \Omega(\CQ)\vert \sim P^3$ for $P\to \infty$ would be
a sharp counterexample to the OSV conjecture, and would have other
interesting implications. As discussed at length in Ref.
\cite{Denef:2007vg,Huang:2007sb}, since $\Omega(\CQ)$ is an index it
is conceivable that the exponentially large contributions might
cancel, leaving asymptotics $\log\vert \Omega(\CQ)\vert \sim
\sqrt{-\hat q_0 P^3}$.  Ref. \cite{Denef:2007vg} argued that such
cancellations are unlikely, but left this central   question
unanswered.

 It is interesting to consider this central question in
the light of the present paper. One way to approach this problem is
via the behavior of ``barely polar degeneracies,'' that is, the
coefficients $\Omega_\delta$ for $\delta$ of order $1$ or smaller
(compared to $P^3$). The entropy enigma suggests that these barely
polar degeneracies grow like $\exp[k P^3]$ as $P\to \infty$ for some
constant $k$.  We are thus led to ask what constraints are imposed
by modular invariance on polar degeneracies, and whether the
existence of terms with large poles $\sim q^{-P^3/24}$ implies,
through anomaly cancellation, that the coefficients of terms with
small or order one poles $\sim q^{-1/\vert P \vert}, \cdots,
q^{-1},\cdots, q^{-2},\dots$   are large. It is convenient to apply
the anomaly cancellation condition in the form
(\ref{eq:cuspanomcan}). The Fourier coefficients $h(n)$ of cusp forms
  (for $\Gamma$, with trivial multiplier system) of weight $k$ grow as
$n^{k/2}$. Although modular invariance therefore bounds the growth of
the polar degeneracies, a lot of freedom remains for these
degeneracies. From these heuristic arguments, it is clear that we must
look elsewhere for an explanation of exponentially large barely polar
degeneracies.

In the following we will refine a   suggestion made  in Ref.
\cite{Denef:2007vg}, p. 117. We make a toy model of the polar terms
of the $(0,4)$ elliptic genus by considering a modular form for
$\Gamma$ with trivial multiplier system (for symplicity) and
considering the polar terms of the negative weight form  $\Phi
\eta^{-\chi}$ where $\chi = P^3 + c_2(X)\cdot P$ and $\Phi$ is a
nonsingular modular form for $\Gamma$ of positive weight
$w_\Phi=\half \chi-1-\half b_2$. As we remarked above, the leading
coefficient $H_{\mu=0}(0)$ is, up to a sign, $I_P \sim P^3/6$ and
therefore in our    toy model $\Phi$ will have a nonzero Petersson
inner product with the Eisenstein series.

To begin, let us sharpen the comments made in \cite{Denef:2007vg}
about the barely polar degeneracies of $\eta^{-\chi}$ for large
$\chi$. For simplicity we assume $\chi$ is a positive integer
divisible by $24$. Let us define  Fourier coefficients by
\begin{equation}
\eta^{-\chi}(\tau) = q^{-\chi/24}\sum_{n=0}^\infty p_\chi(n) q^n.
\end{equation}
We are considering degeneracies for $n= \frac{\chi}{24} + \ell$ with
$\ell $ fixed as $\chi \to \infty$ (and of either sign) so the usual
Hardy-Ramanujan analysis (``Cardy formula'')  is slightly altered. A
naive saddle-point analysis proceeds by writing
\begin{equation}
p_\chi(n)= \int_{\tau_0}^{\tau_0+1}  e^{-2\pi i
(n-\chi/24)\tau} \frac{1}{\eta^\chi} d\tau \cong
\int_{\tau_0}^{\tau_0+1} e^{-2\pi i (n-\chi/24)\tau +
\frac{\chi}{2}\log(-i \tau) + \frac{i\pi \chi}{12 \tau}}d\tau.
\end{equation}
In contrast to the usual estimate, it is now the second and third terms in the
exponential which dominate the saddle point. In this way we estimate
\begin{equation}
p_\chi\left(\frac{\chi}{24} + \ell\right) \sim_{\chi\to \infty} const.
\chi^{-1/2} \exp\left( \frac{\chi}{2} \left(1+ \log\frac{\pi}{6}\right) +
\frac{\pi^2}{3} \ell\right). \end{equation}
This agrees very well with a numerical analysis of $\log
p_\chi(\chi/24)$ in Ref. \cite{Denef:2007vg} (p.117). Moreover, we
see that although the degeneracies grow exponentially with $\ell$,
the proportionality between $p_\chi\left(\frac{\chi}{24} +
\ell\right)$ and $p_\chi\left(\frac{\chi}{24} + \ell + 1\right)$ is
not exponential in $\chi$. This agrees with the earlier statement
that the anomaly cancellation bounds the growth of the polar
degeneracies.

It is interesting to compare with the Rademacher formula for
$p_\chi(\chi/24)$:
\begin{equation}
p_\chi\left(\frac{\chi}{24}\right) = 2\pi \sum_{0 \leq n < \frac{\chi}{24}}
p_\chi(n) \frac{\left(2\pi \vert n - \frac{\chi}{24}\vert\right)^{1+
\chi/2}}{\Gamma(2+\chi/2)} \sum_{c=1}^\infty c^{-2 - \chi/2}
K_c\left(0,n-\frac{\chi}{24}\right).
\end{equation}
We can use a beautiful formula of Ramanujan: \footnote{To show this we
  first relate the relevant Kloosterman sum to the M\"obius
  function $\mu(n)$: $\sum_{{a=1 \atop (a,c)=1}}^c
  e(n\frac{a}{c})=\sum_{m|(c,n)}\mu(\frac{c}{m})m$ (page 160 of Ref. \cite{apostol:1976}).
  We substitute this   identity in the left hand side of
  Eqn. (\ref{eq:ramanujan}). Application of $\zeta(s)\sum_{n=1}^\infty
  \frac{\mu(n)}{n^s}=1$ leads then to the claimed identity.}
\begin{equation}
\label{eq:ramanujan}
\sum_{c=1}^\infty c^{-s} K_c(0,n) = \frac{\sigma_{1-s}(n)}{\zeta(s)},
\end{equation}
to simplify our formula to:
\begin{equation}
p_\chi\left(\frac{\chi}{24}\right) = 2\pi \sum_{0 \leq n < \frac{\chi}{24}}
p_\chi(n) \frac{\left(2\pi \vert n - \frac{\chi}{24}\vert\right)^{1+
\chi/2}}{\Gamma(2+\chi/2)}
\frac{\sigma_{-1-\chi/2}(\frac{\chi}{24}-n)}{\zeta(2+\chi/2)}.
\end{equation}
Now, note for large $\chi$ there is a very large denominator from
the Gamma function. The factor $\left(2\pi \vert n -
\frac{\chi}{24}\vert\right)^{1+ \chi/2}$ starts very large for $n=0$
and falls exponentially rapidly. Meanwhile, notice that since the
index on the divisor sum is negative the factor
$\sigma_{-1-\chi/2}(\frac{\chi}{24}-n)$ is a slowly varying function
of $n$, and strictly smaller than $\frac{\chi}{24}-n$. Thus, the sum
is dominated by the terms $n=0$. Using Stirling's formula we find
that the contribution of the $n=0$ term is
\begin{equation}
\sigma_{-1-\chi/2}\left(\frac{\chi}{24}\right) \times const. \times
\chi^{-1/2}\exp\left(\frac{\chi}{2} \left(1+ \log\frac{\pi}{6}\right)\right).
\end{equation}
in agreement with the naive evaluation. Thus we learn that the
contribution of the \textit{extreme polar states} in the Rademacher
expansion gives the dominant contribution to the constant term.

Now let us turn to the numerator $\Phi$.  A similar discussion
applies to the contributions of $\Phi$ to the barely polar
degeneracies. If $\Phi$ is a nonsingular modular form of weight $w$
with $\Phi(\tau) = \sum_{n\geq 0} \hat \phi(n) q^n$ then a naive
saddle point evaluation of the Fourier coefficients $\hat\phi(n)$
gives
\begin{equation}
\hat \phi(n) \sim \pm \frac{\hat \phi(0)}{\sqrt{2\pi}} w^{-w+\half}
e^{w \left(1+ \log(2\pi)\right)} n^{w-1}\left(1 +
\CO(e^{-4\pi^2n/w})\right)
\end{equation}
(Although this is naive, numerical checks indicate it is valid.)  To
estimate the biggest contribution of the Fourier coefficients of
$\Phi$  to the constant term in $\eta^{-\chi}\Phi$  we apply this to
$w=w_\Phi= \half \chi-\half b_2 -1$ and $n = \frac{\chi}{24} $
yielding, remarkably,
\be const. \chi^{-1/2} \exp\left[ \frac{\chi}{2}\left(1+ \log
\frac{\pi}{6}\right)\right] \ee
having the same order of exponential growth as the barely polar
terms of $\eta^{-\chi}$.  Thus, in our model for polar degeneracies
the barely polar degeneracies are indeed expected to grow
exponentially in $\chi$.

It is conceivable that this kind of estimate could be rigorously
applied to estimate the coefficients near the cosmic censorship
bound in the $(0,4)$ elliptic genus, and it would be very
interesting to do so.

\subsection{Enumerative geometry}
\label{subsec:dtcuspform}

As a final application of the fareytail expansion, we would like to
point out its potential relevance to problems in enumerative
geometry. The Fourier coefficients $\Omega_\delta $ of the
$\mathcal{N}=(0,4)$ elliptic genera are the degeneracies of bound states of D4-, D2-,
and D0-branes on a Calabi-Yau manifold $X$. From a more mathematical
perspective, these are (generalized) Donaldson-Thomas invariants,
which count the stable coherent sheaves on $X$ with given Chern classes. 
The BPS degeneracies (or equivalently Donaldson-Thomas invariants) are subject to wall-crossing
behavior, since the BPS-states are not stable for all values of the (complexified) K\"ahler moduli $t$ (specified at spatial 
infinity in the black hole solution). The complexified K\"ahler
moduli are given by $t=B+iJ$, where $B$ is the anti-symmetric tensor
field and $J$ is the K\"ahler class. The generating function of the
BPS-degeneracies has only an interpretation as an $\mathcal{N}=(0,4)$ elliptic genus
\cite{Maldacena:1997de, Denef:2007vg} in the large K\"ahler
limit. Ref. \cite{deBoer:2008fk} argues more precisely that the
$(0,4)$ SCFT analysis is only valid if the $t^a$ are chosen such that
$t^a=d^{ab}q_b+i\lambda p^a$ with $\lambda \to \infty$. 

The fact that a class of DT-invariants are enumerated by a modular
form has interesting consequences. For example, section
\ref{sec:anomalies} discussed how a modular anomaly arises if the
polar coefficients do not satisfy  certain constraints. These
constraints are such that a linear combination of cuspidal Poincar\'e
series vanishes. The constraints are given in the form 
\be \label{eq:cuspanomcan2} \forall j\qquad  \sum_{ \delta>0 }
\Omega_\delta \, \overline{h^j(\lfloor \delta \rfloor)}=0 , \ee
where the $h^j(n)$ are Fourier coefficients of an orthonormal basis of
cusp forms. Therefore, we see that interesting relations exist among the
coefficients of cusp forms and DT-invariants in a specific chamber of
the moduli space. Generically, it is very difficult to find such
relations among cusp forms. A concrete example where this phenomenon
occurs, is the case where the M5-brane wraps the hyperplane section of
the bicubic in $\mathbb{CP}^5$. Ref. \cite{Gaiotto:2007cd} computes
explicitly the elliptic genus of this configuration (and several others) by a
determination of the polar degeneracies using algebraic geometry and
Gromov-Witten invariants. Interestingly, a relation among the polar
coefficients was found, which was explained in
Ref. \cite{Manschot:2008zb} as a consequence of the existence of a (vector-valued) cusp form
with the relevant properties.

In some respects, the $(0,4)$ elliptic genus can be seen as a
generalization of the partition function of bound states of D4-D2-D0 branes on K3. For example, if the 11-dimensional geometry is chosen to be  
$\mathbb{R}^{5}\times T^2\times$K3, and a single M5-brane wraps
$T^2\times$K3, then the $\mathcal{N}=(0,4)$ elliptic genus becomes
\be
\label{eq:egK3}
\chi(\tau, z)_{\mathrm{K3}}=\frac{\Theta_{\Gamma_{3,19}}(\tau, \bar
  \tau, z)}{\eta(\tau)^{24}}.
\ee
Note that since this geometry preserves more supersymmetry a factor
$F^4$ needs to be inserted in the trace (\ref{eq:04eg}), instead of
$F^2$. $\Gamma_{3,19}$ is the lattice of the second cohomology of
K3. We observe that $\eta(\tau)^{-24}$ provides us the number of
BPS-degeneracies of D0-branes as well as D2-branes on K3. The
D0-branes are the physical equivalent of the Hilbert scheme of
points. This partition function is earlier computed from this perspective in 
\cite{gottsche:1990}. Recently, the interpretation of $\eta(\tau)^{-24}$  as a
generating function for D2-branes wrapping cycles in K3 has been put
on a firmer mathematical basis \cite{klemm:2008}. It provides the (reduced) Gromov-Witten
invariants of K3. The (0,4) elliptic genus in the case of a proper
Calabi-Yau threefold $X$ and possibly multiple M5-branes, is a
major generalization of (\ref{eq:egK3}). We expect that it can play an
important role in problems of enumerative geometry related to
Calabi-Yau threefolds.


\section{Non-Holomorphic Partition Functions}
\label{sec:nonholomorphic}\setcounter{equation}{0}

This section explains how the anomalous transformation property of $\hat
S_{\mathrm{Reg}}^{(\delta)}(\tau)$ under $\Gamma$ in Eq.
(\ref{eq:truetransform}), can be corrected by the addition of a
non-holomorphic term to produce a covariant object. Section
\ref{sec:anomalies} shows that a proper choice of polar degeneracies
can result in the vanishing of the shift in Eqs.
(\ref{eq:truetransform}) or (\ref{eq:transformvectorform}). However,
physics might prescribe a set of polar degeneracies which can not be
consistently  extended to a holomorphic modular form with the
required transformation properties. Holomorphy is useful, but
diffeomorphism invariance is fundamental, hence in   such a
situation there is necessarily a holomorphic anomaly. We now explore
what can be said about such holomorphic anomalies from the viewpoint
of this paper.

Eq. (\ref{eq:transformp}) shows that if we add a non-holomorphic term as
in
\be \tilde S^{(\delta)}_{\rm Reg}(\tau,\bar \tau)=\hat
S^{(\delta)}_{\rm
  Reg}(\tau) - p(\tau, \bar \tau, \overline{G^{(\delta)}})
\ee
then the new function $\tilde S^{(\delta)}_{\rm Reg}(\tau,\bar
\tau)$ transforms covariantly.  In this way we can trade the modular
anomaly for a holomorphic anomaly.  To study its properties more
precisely, we rewrite $p(\tau, \bar \tau, \overline{G^{(\delta)}})$
as
\be
\frac{1}{\Gamma(1-w)}\int_{\bar \tau}^{-i\infty}
\overline{G^{(\delta)}(z)} (\bar z- \tau)^{-w} d\bar z =
\frac{(-2i\tau_2)^{1-w}}{\Gamma(1-w)}\int_1^\infty
\overline{G^{(\delta)}(\bar \tau + 2ui\tau_2)}u^{-w} du. \ee
{}From the first expression it is clear that    $\tilde
S^{(\delta)}_{\rm Reg}(\tau,\bar \tau)$ satisfies the  holomorphic
anomaly equation
\be \label{eq:holanomaly} \frac{\partial }{\partial \bar \tau}
\tilde S^{(\delta)}_{\rm
  Reg}(\tau,\bar \tau)=\frac{(-2i\tau_2)^{-w}}{\Gamma(1-w)}\overline{
  G^{(\delta)}(\tau)}.
\ee
\noi Of course, such a non-holomorphic correction is far from being
unique! The above choice is distinguished by the fact that $\tilde
S^{(\delta)}_{\rm Reg}(\tau,\bar \tau)$ is annihilated by a
Laplacian given by $\Delta=\frac{\partial}{\partial
  \tau} \tau_2^w \frac{\partial}{\partial \bar \tau}$.
Note that it also   reduces to a polynomial in $\tau$ for $-w \in
\mathbb{N}$.

The holomorphic anomaly described here is similar to the one appearing
  for the $w=\frac{3}{2}$ modular forms discussed in \cite{zagier:1975,
  hirzebruch:1976}. In physics, such
   holomorphic anomalies arise   in the
partition function of $\mathcal{N}=4$ topologically twisted
Yang-Mills theory on $\mathbb{CP}^2$ with gauge group $SO(3)$
\cite{Vafa:1994tf}, and also in the context of Donaldson invariants
\cite{Moore:1997pc}. Now, as reviewed in section \ref{sec:review},
if we consider an M5-brane partition function on $\Sigma \times T^2$
then for small $T^2$ we would expect the partition function to be
related to the four-dimensional gauge theory computations of
\cite{Vafa:1994tf}. On the other hand in the limit when the K\"ahler
class of the $T^2$ is much larger than those of $\Sigma$, and
$\Sigma$ is embedded in a Calabi-Yau manifold,  a $(0,4)$ conformal
field theory analysis analogous to that of \cite{Maldacena:1997de}
should be applicable. This suggests that there might be holomorphic
anomalies in the $(0,4)$ elliptic genus. \footnote{Exactly this
suggestion has been made previously by D. Gaiotto in a seminar at
Princeton, Oct. 13 2006.}

As a possible example of this situation consider wrapping  an
M5-brane on a rigid divisor equal to $\mathbb{CP}^2$ in a suitable
Calabi-Yau (e.g. the Calabi-Yau elliptic fibration over
$\mathbb{CP}^2$). Ref. \cite{Vafa:1994tf} calculates the partition
function of the twisted gauge theory. The coefficients of this
partition function are the Euler numbers of the moduli space of
instantons. In the case of  $\mathbb{CP}^2$ with gauge group $SO(3)$
Ref. \cite{Vafa:1994tf} gives two partition functions, $Z_0(\tau,
\bar \tau)$ and $Z_1(\tau, \bar \tau)$, related to the two different
possibilities for the second Stiefel-Whitney class $w_2$ of $SO(3)$ bundles on
$\mathbb{CP}^2$. $Z_0(\tau, \bar \tau)$ and $Z_1(\tau, \bar \tau)$
transform as a modular vector under $\Gamma$. The holomorphic
anomaly for $Z_\mu(\tau, \bar \tau)$, given in Ref.
\cite{Vafa:1994tf}, is 
\be \label{eq:holanomaly2} \frac{\partial}{\partial \bar \tau}
Z_\mu(\tau,\bar \tau)=\frac{3}{16\pi
i\tau_2^{3/2}}\frac{1}{\eta(\tau)^6} \sum_{n\in \mathbb{Z}+\frac{\mu}{2}}\bar q^{n^2}=\frac{3}{16\pi
i\tau_2^{3/2}}\frac{1}{\eta(\tau)^6} \overline{ \theta_{3-\mu}(2\tau)}, \ee
where $\theta_{3-\mu}(\tau)$ are the standard Jacobi theta
functions. From this one can derive the modular transformations of the
purely holomorphic partition function:
\be
\label{eq:vwtransform}
Z_\mu(\gamma(\tau))=j(\gamma,\tau)^{-\frac{3}{2}}M(\gamma)_\mu^\nu\left[Z_\nu(\tau)+\frac{3 e(-\frac{1}{8})}{2\sqrt{2\pi}\,\eta(\tau)^6}
p\left(\tau,\gamma^{-1}(-i\infty),\overline{ \theta_{3-\nu}(2\,\cdot )}\right)\right],
\ee
where $M(\gamma)$ is the multiplier system generated from
\be \label{gaugemultsys}
M(T) =  \begin{pmatrix} e(-1/4) & 0 \\ 0 & -1 \\
\end{pmatrix}, \qquad M(S) = e(-1/8) \frac{1}{\sqrt{2}}\begin{pmatrix} 1 & 1 \\ 1
& -1 \\ \end{pmatrix}. \ee

 To compare these partition functions with a dual supergravity
partition function we must recall that the gauge theory dual to the
string theory will include singleton degrees of freedom leading to
extra $U(1)$ factors in the gauge group. (See
\cite{Maldacena:2001ss}, appendix B, or \cite{Gukov:2004id,
Belov:2004ht}.) In the present case we should presumably compare to
a theory with gauge group   $U(2)$.   After inclusion
of the $U(1)$ degrees of freedom, we obtain
\be
\label{eq:cp2vw}
\chi(\tau,\bar \tau, \bar z)=Z_0(\tau, \bar \tau)\overline{\theta_2(2\tau,2z)}
-Z_1(\tau, \bar \tau)\overline{\theta_3(2\tau,2z)}.
\ee
$\chi(\tau,\bar \tau,\bar z)$ transforms under $\Gamma$ with weight
$(-\frac{3}{2},\frac{1}{2})$ and multiplier system.  This clearly
resembles an elliptic genus of a $(0,4)$ SCFT as given in Eq. (\ref{singlem5}).

Let us therefore contrast these formulae with what would be expected from the
viewpoint of this paper. We might expect to be able to construct the
partition function -- in the AdS$_3$ regime
-- from a Poincar\'e series based on its polar part. A priori, this
partition function does not need to equal $\chi(\tau,\bar \tau, \bar z)$
since we might not be able to rely on modular invariance and/or holomorphy.  Therefore, we
distinguish the fareytail partition function and denote it by
$\chi^{\mathrm{FT}}(\tau,\bar \tau, \bar z)$. The theta functions in
Eq. (\ref{eq:cp2vw}) can be derived from this point of view as a specialization of
Eq. (\ref{eq:siegelnarain}). Note that $\mu^\|$ is $0$ when the
second Stiefel-Whitney class $w_2$ of the $SO(3)$ bundle is trivial,
and equal to 1 when $w_2$ is non-trivial. 

The comparison reduces now to a comparison of the holomorphic part
of $Z_\mu(\tau,\bar \tau)$, $Z_\mu(\tau)$, with the vector-valued
modular form constructed by the Poincar\'e series. We label the
constructed vector-valued modular form by ``FT'':
$Z_\mu^{\mathrm{FT}}(\tau)$. The polar part of
$Z_\mu^{\mathrm{FT}}(\tau)$ is equal to the polar part of
$Z_\mu(\tau)$, if we assume that the polar part is not renormalized
as we continue to the AdS$_3$ regime. $Z_0(\tau)$ has a polar term
equal to $-\frac{1}{4}q^{-\frac{1}{4}}$ while $Z_1( \tau)$ does not
contain a polar term. Therefore, we attempt to construct with the
fareytail a modular form of weight $-3/2$, with multiplier system
given by Eq. (\ref{gaugemultsys}) and polar term given by $\delta =
\begin{pmatrix}\frac{1}{4}\\ 0 \\ \end{pmatrix}$.
The obstruction to the construction of a holomorphic modular form with
these properties is given by a space of vector-valued cusp forms as
discussed extensively in previous sections. The space of these cusp
forms turns out to be non-vanishing in this case. A vector-valued cusp
form of weight $7/2$ and the appropriate multiplier system is given by
\be
\eta(\tau)^6\left(\begin{array}{c} \theta_{3}(2\tau)  \\ \theta_{2}(2\tau) \end{array} \right).
\ee
Using the dimension formulas for vector-valued modular forms, one can show that this form is the 
unique cusp form with the required properties. See Ref. \cite{Manschot:2008zb} for more details and illustrations of dimension 
formulas. Then we find the following transformation law for   
$Z_\mu^{\mathrm{FT}}(\tau)$
\be
\label{eq:fttransform}
Z_\mu^{\mathrm{FT}}(\gamma(\tau))=j(\gamma,\tau)^{-\frac{3}{2}}M(\gamma)_\mu^\nu\left[Z_\nu^{\mathrm{FT}}(\tau)+\frac{1}{4}
p\left(\tau,\gamma^{-1}(-i\infty),\overline{\eta^6 \, \theta_{3-\nu}(2\,\cdot )}\right)\right].
\ee
The factor $\frac{1}{4}$ in front of the period function is a
consequence of the coefficient of the polar term.

A simple check whether the fareytail can reproduce the gauge theory
partition function is a comparison of the anomalies under modular
transformations. Even without a detailed analysis, we can observe
qualitative differences between the shifts. An important difference
is the behavior for $\im(\tau)\to \infty$. In this limit the shift
in Eq. (\ref{eq:vwtransform}) grows   exponentially whereas the
period function in Eq. (\ref{eq:fttransform}) vanishes.
 This shows clearly that the holomorphic fareytail does not
equal the generating function of the Euler numbers of instanton moduli spaces.

As a consequence of the different modular anomalies, the associated holomorphic anomalies are different.
The holomorphic anomaly given by Eq. (\ref{eq:holanomaly2}) is not annihilated by the Laplacian $\Delta$.
 Another difference is that for $\im(\tau)\to \infty$, the right hand side of Eq. (\ref{eq:holanomaly2}) grows
exponentially (for $\mu=0$).

This raises the question of what the elliptic genus of the
$\mathcal{N}=(0,4)$ SCFT on the boundary of AdS$_3$ really is. The
results of this section are clearly inconclusive. We are considering
several possible resolutions and we hope to address them in future
work.

\section{Conclusion}
\label{sec:conclusion}\setcounter{equation}{0}

In this paper we have revisited the ``fareytail expansion'' of
\cite{Dijkgraaf:2000fq}, and have improved on the story in many
ways. We have shown how to regularize the relevant Poincar\'e series
so that we have an expansion for the partition function, and not its
 ``fareytail transform.'' The latter is problematic, and now rendered
 irrelevant.

The modern fareytail is well-suited to the earlier applications of
fareytail expansions. It is relevant for
the program of determining the black hole entropy by study of the
near horizon microstates. We have argued that the new expansion is
consistent with the OSV conjecture at strong topological string
coupling.

In addition, the modern fareytail contains a number of interesting
new aspects. This includes new wrinkles on the interpretation of the
expansion in the AdS/CFT context, as well as new corrections to the
OSV formula at weak coupling.  Moreover, we have given an extended
discussion how the regularization can give rise to a modular or
holomorphic anomaly. The modular anomalies can be described
  in terms of period functions of positive weight cusp forms. The
  holomorphic anomaly is compared with a similar anomaly
  appearing in the partition function of  $\mathcal{N}=4$ Yang-Mills on
  $\mathbb{CP}^2$.

There are further implications of the new fareytail, not discussed
in this paper, which might prove fruitful  for future study. One of
these questions concerns the spaces of obstructions to the
construction of the modular forms. We would like to sharpen our
understanding by computing, for example, the precise dimension of
the space of obstructions. Another point which deserves further
study is the possibility of holomorphic anomalies in the elliptic
genus.  A better understanding of the relation of the holomorphic
anomalies to those of topological $\mathcal{N}=4$ Yang-Mills  is
desirable.

Finally, we mention a more speculative connection to arithmetic
varieties. Arithmetic varieties appeared earlier in the context of
black holes in Ref. \cite{Moore:2004fg,Moore:1998zu,Moore:1998pn}.
It is possible to associate arithmetic varieties in two distinct
ways to a polar term. On the one hand, a  polar term corresponds to
several split attractor flows \cite{Denef:2007vg}. The split
attractor flows of Denef end on regular attractor points. The
conjectures in Ref. \cite{Moore:2004fg,Moore:1998zu,Moore:1998pn}
state that the Calabi-Yau at a regular attractor point is an
arithmetic variety. On the other hand, arithmetic varieties can also
appear in an alternative way via the cusp form which is associated
to the polar term. The cusp form can be decomposed into Hecke
eigenforms. The Hecke eigenforms can be related to
arithmetic Calabi-Yau manifolds (usually with dimension larger than 3), generalizing the celebrated case of
the elliptic curve. For a review see for example Ref.
\cite{Yui:2003}. Thus we have two different ways to relate a polar
term to an arithmetic manifold. It would be quite interesting if
this correspondence turns out to have any arithmetic significance.

\bigskip
\begin{center} {\bf Acknowledgments}\end{center}
J. M. would like to thank the Department of Physics \& Astronomy of
Rutgers University and the School of Natural Sciences of the
Institute for Advanced Study  for hospitality during the completion
of this work. He is grateful to E. Verlinde for discussions and
encouragement to survey the problems related to the fareytail
transform.

G. M. would like to thank H. Ooguri, whose crucial question led to
the discovery of the problems with   the fareytail transform. He
also thanks D. Zagier for important discussions and correspondence
and he thanks   S. Gukov, H. Ooguri, and  C. Vafa, for collaboration
on closely related matters.

In addition, we thank J. de Boer, F. Denef, G. van der Geer, E. Diaconescu, J.
Maldacena, S. Miller, P. Sarnak, B. van Rees and E. Witten for
helpful discussions. The research of J. M. is supported by the
Foundation of Fundamental Research on Matter (FOM). The work of G.
M. is supported by the US DOE under grant DE-FG02-96ER40949.

\appendix
\section{Technicalities of the Modern Fareytail}
\label{sec:derivation}\setcounter{equation}{0}
\subsection{Derivation}
This appendix derives Eq. (\ref{eq:sumcd}). The derivation is in
some sense a reversed version of the analysis in Ref.
\cite{niebur1974}. We start with a vector-valued modular, and derive
Eq. (\ref{eq:sumcd}) based on its Fourier coefficients, which are
calculated by the Rademacher circle method. Whereas Ref.
\cite{niebur1974} basically starts at the other end, and determines
its Fourier coefficients together with its transformation
properties. We take the opportunity to generalize the result to
vector-valued modular forms.

To start, we state the transformation properties of a vector-valued modular form
\be
\label{eq:modularvector2}
f_\mu(\gamma(\tau))= M(\gamma)^\nu_\mu(c\tau+d)^w f_\nu(\tau).
\ee

\noi with $\gamma=\left(\begin{array}{cc} a & b
  \\ c & d\end{array} \right)\in \Gamma$. We take $w\leq 0$ and use $-\pi
  < \arg(z)\leq \pi$ as domain for the argument of a complex variable
  $z$. The Fourier expansion of  the modular vector is given by
\be
f_\mu(\tau)=\sum_{m=0}^\infty F_\mu(m) q^{m-\Delta_\mu},
\ee

\noi where $F_\mu(0)\neq 0$ is the lowest non-zero coefficient. The part of $f_\mu(\tau)$
with $m-\Delta_\mu<0$ is denoted as its polar part
$f^-_\mu(\tau)$, because of the divergence of these
terms when $\tau \to i\infty$. The series with $m-\Delta_\mu\geq 0$ is
correspondingly called the non-polar part, $f^+_\mu(\tau)$. Note that
for transformations $\gamma_n(\tau)=\tau+n$, $M(\gamma)_\mu^\nu$ is given by
  $\delta_\mu^\nu e(-\Delta_\mu n)$. The Fourier coefficients (with
  $m-\Delta_\mu\geq 0$) are determined by the Rademacher circle method or Farey
  fractions \cite{rademacher1938:1}. This method is beautifully applied
  to $1/\eta(\tau)$ in  Ref. \cite{apostol} and generalized to vector-valued
  modular forms in  Ref. \cite{Dijkgraaf:2000fq}. The Fourier  coefficients are
  given by the infinite series
\begin{eqnarray}
\label{eq:fourier}
F_\mu(m)&=&2\pi \sum_{n-\Delta_\nu<0} F_\nu (n)\sum_{c=1}^\infty
\frac{1}{c} K_c(m-\Delta_\mu,n-\Delta_\nu) \\
&&\times \left(\frac{|n-\Delta_\nu|}{m-\Delta_\mu}
\right)^{(1-w)/2}I_{1-w}\left(\frac{4\pi}{c}\sqrt{(m-\Delta_\mu)|n-\Delta_\nu|}
\right), \nonumber
\end{eqnarray}

\noi where $I_{\nu}(z)$ is the modified Bessel function of the first
kind. $I_{\nu}(z)$ is given as an infinite sum by
\be
\label{eq:besselsum}
I_\nu(z)=\left(\frac{z}{2}\right)^\nu \sum_{k=0}^\infty \frac{\left(\frac{1}{4}z^2\right)^k}{k!\Gamma(\nu+k+1)}.
\ee

\noi $K_c(m-\Delta_\mu, n-\Delta_\nu)$ is a generalized version of
the Kloosterman sum
 \be \label{eq:kloosterman}
K_c(m-\Delta_\mu,n-\Delta_\nu):=i^{-w}\sum_{{ -c \leq d<0} \atop
{(c,d)=1}} M^{-1}(\gamma)_\mu^\nu e\left((n-\Delta_\nu)
\frac{a}{c}+(m-\Delta_\mu)\frac{d}{c}\right), \ee

\noi with $\gamma=\left(\begin{array}{cc} a & b \\ c & d\end{array}
  \right)\in \Gamma$, thus $ad= 1 \mod c$. We have taken a specific
  domain for $d$ in the Kloosterman sum. This is necessary since $\Delta_\mu$ is in
general not an integer. The dependence on $a$ in the exponent and in
  $M^{-1}(\gamma)_\mu^\nu$  via $\gamma$ combine such that the
  product with the
  generalized Kloosterman sum is independent of $a$. The factor of
  $i^{-w}$ in front of the sum is a consequence of the definition of
  $M(\gamma)_\mu^\nu$ in Eq. (\ref{eq:modularvector2}). Finally, if
  $m-\Delta_\mu=0$ we should take a limit as $m-\Delta_\mu\to 0$.

Since $M(\gamma)$ is unitary, the generalized Kloosterman sum is
bounded above by the Euler totient function $\phi(c)\leq
c$. For later use, we need an estimate of the generalized Kloosterman
sum. Weil has derived a particularly strong bound for $K_c(m,n)$ when $m,n \in
\mathbb{Z}$ and a trivial multiplier system. He estimated that
$K_c(m,n)$ is bounded above by
$\mathcal{O}(c^{\frac{1}{2}+\epsilon})$. We do not need such a strong
bound. For our applications with $w<0$, the upperbound of the
Kloosterman sum by $c$ suffices. For the example in the introduction
with $w=0$ and a trivial multiplier system (Eq. (\ref{eq:reggravpath})), an
estimate $c^{1-\epsilon}$ with $\epsilon>0$ is necessary. Such a bound
can be established in an elementary way, see for example \cite{heath-brown:2000}. We
do not attempt to establish a non-trivial bound for Kloosterman sums
arising from modular forms with $w=0$ and a non-trivial multiplier
system. 

Our strategy to derive Eq. (\ref{eq:sumcd}) is fairly
straightforward. We substitute the expression for the Fourier
coefficients in the Fourier series for the non-polar part of
$f_\mu(\tau)$. Then we use the formulas given in appendices
\ref{sec:period} and \ref{sec:lipschitz} to rewrite $f_\mu(\tau)$ in
the form of  Eq. (\ref{eq:sumcd}). After  the substitution of the
Fourier coefficients Eq. (\ref{eq:fourier}) and Kloosterman sum Eq.
(\ref{eq:kloosterman}), we insert the series expansion of the Bessel
function Eq. (\ref{eq:besselsum}). We obtain
\begin{eqnarray}
\label{eq:insertbessel}
f^+_\mu(\tau)&=&\sum_{m-\Delta_\mu \geq 0}F_\mu(m)q^{m-\Delta_\mu}\\
&=&\sum_{n-\Delta_\nu<0}\sum_{c=1}^\infty \sum_{{-c \leq d<0} \atop {(c,d)=1}}
\sum_{k=0}^\infty i^{-w} M^{-1}(\gamma)_\mu^\nu F_\nu(n)\left(\frac{2\pi}{c} \right)^{2k+2-w} \frac{|n-\Delta_\nu|^{k+1-w}}{\Gamma(k+2-w)} \nonumber \\
&&\times e\left((n-\Delta_\nu)\frac{a}{c}\right)
\sum_{m-\Delta_\mu \geq 0}
\frac{(m-\Delta_\mu)^k}{k!}e\left((m-\Delta_\mu)\left(\tau+\frac{d}{c}\right)\right),
\nonumber
\end{eqnarray}

\noi where we interchanged the sum over $m$ with the other four sums
and grouped the terms dependent on $m$. We apply the
Lipschitz summation formula (\ref{eq:lipschitz}) to the sum over
$m$, the new summation variable will be denoted by $l$. The error term $E(\tau,k+1,N+\frac{1}{2})$
vanishes in the limit $N\to \infty$, except when $k=0$ and
$\Delta_\mu\in \mathbb{N}$. When the error term does not vanish, we
get an additional constant. This constant is equal to
$\frac{1}{2}F_\mu(\Delta_\mu)$ and is given by
\begin{eqnarray}
\label{eq:constant}
\frac{1}{2}F_\mu(\Delta_\mu)&&=  \\
&&\left\{ \begin{array}{cc} \pi \sum_{n-\Delta_\nu<0}\frac{(2\pi
  |n-\Delta_\nu|)^{1-w}}{\Gamma(2-w)}F_\nu(n)\sum_{c=1}^\infty
  c^{w-2}K_c(0_\mu,n-\Delta_\nu), & \Delta_\mu \in \mathbb{N},\\ 0, &
  \Delta_\mu \not\in \mathbb{N}, \end{array} \right. \nonumber
\end{eqnarray}

\noi where $0_\mu$ is a vector all of whose components are zero. The
fact that the right hand side of Eq. (\ref{eq:constant}) is equal to
$\frac{1}{2}F_\mu(\Delta_\mu)$ can be shown for example by
Eq. (\ref{eq:fourier}) for $F_\mu(\Delta_\mu)$ and the limiting
behavior of the Bessel function for $z\to 0$: $\lim_{z\to
  0}I_\nu(z)=\left(\frac{z}{2}\right)^{\nu}\frac{1}{\Gamma(\nu+1)}$. We
get after interchanging the sum over $k$ and $l$
\begin{eqnarray}
\label{eq:subslipschitz}
f_\mu^+(\tau)&=&\frac{1}{2}F_\mu(\Delta_\mu)+ \sum_{n-\Delta_\nu<0}
  \sum_{c=1}^{\infty} \sum_{{-c \leq d <0} \atop (c,d)=1} \lim_{N\to \infty}   \sum_{l=-N}^N M^{-1}(\gamma)_\mu^\nu F_\nu(n)
  e((n-\Delta_\nu)\frac{a}{c}) \nonumber \\
&&\times \frac{1}{(c\tau+d+cl)^w} e(\Delta_\mu l)
 \sum_{k=0}^\infty \frac{1}{\Gamma(k+2-w)} \left(\frac{2\pi i |n-\Delta_\nu|}{c(c\tau+d+cl)}\right)^{k+1-w}.
\end{eqnarray}

\noi The exchange of the sum over $k$ and $l$ is allowed because the
sums are absolutely convergent for $k>0$. In case $k=0$, the sum over $l$ in the
limit $N\to \infty$ is as well convergent. This is shown using the
weak bound on the Kloosterman sum, to which we referred earlier.

The sums over $c$ and $d$ can be such that they have an equal
upperbound. This is clear for $k>0$, but to show it for $k=0$ is
slightly subtle. First, we incorporate the sum over $l$ in the sum
over $d$. Since the sum over $l$ and $d$ is convergent for finite $c$,
we can choose for $|d|$ an upperbound $N$ for which we take the limit
$N\to \infty$. We thus get a sum of the form
\be
\label{eq:sumN}
\sum_{c=1}^\infty\lim_{N\to \infty} \sum_{{|d|\leq N \atop (c,d)=1}}M^{-1}(\gamma)_\mu^\nu\frac{e((n-\Delta_\nu)\frac{a}{c})}{c^{1-w}(c\tau+d)}
\ee

\noi Where we used that $e(\Delta_\mu
l)\delta_\mu^\nu=M^{-1}(\gamma_l)_\mu^\nu$ and Eq. (\ref{eq:g1g2}) to
include $e(\Delta_\mu l)$ in
$M^{-1}(\gamma)_\mu^\nu$. Ref. \cite{rademacher:1939} shows that
\be
\label{eq:sumK}
\lim_{K\to \infty}\sum_{c=1}^K \lim_{N\to \infty}\sum_{{K<|d|\leq N
    \atop (c,d)=1}}M^{-1}(\gamma)_\mu^\nu \frac{e((n-\Delta_\nu)\frac{a}{c})}{c^{1-w}(c\tau+d)}=0,
\ee

\noi in case $M(\gamma)=1$ and $(n-\Delta_\nu)=-1$. We can show in a
similar way that the generalization holds as well. To this end
define the matrix $g(d)_\mu^\nu $ (with $-\delta_\nu=n-\Delta_\nu$)
\be
g(d)^\nu_\mu=\left\{\begin{array}{cc} M^{-1}(\gamma)_\mu^\nu
e(-\delta_\nu\frac{a}{c}), & \mathrm{for}\,\, (c,d)=1, \\ 0, & \mathrm{otherwise}. \end{array} \right.
\ee
Using that $M(\gamma_l)_\mu^\nu=\delta_\mu^\nu e(-\delta_\nu
l)$ (where $\delta_\mu^\nu$ should not be confused with
$\delta_\nu$), we observe that $e(-\delta_\mu\frac{d}{c})g(d)_\mu^\nu$ is
periodic in $d$ modulo $c$. Therefore, $e(-\delta_\mu\frac{d}{c})g(d)_\mu^\nu$
has a Fourier expansion, and we find for $g(d)_\mu^\nu$
\be
g(d)_\mu^\nu=\sum_{j=1}^c \left(B_{j,c}\right)_\mu^\nu e\left((j+\delta_\mu)\frac{d}{c}\right),
\ee
with
\be
\left(B_{j,c}\right)_\mu^\nu=\frac{1}{c}\sum_{{d'=1 \atop (c,d')=1}}^c
M^{-1}(\gamma)_\mu^\nu e\left(-\delta_\nu\frac{a}{c}-(j+\delta_\mu)\frac{d'}{c}\right).
\ee
$B_{j,c}$ contains a Kloosterman sum, and with the bound $c^{1-\epsilon}$  on the vector-valued
Kloos\-terman sums (see the discussion below Eq.
(\ref{eq:kloosterman})), we obtain $\mathcal{O}(c^{-\epsilon})$ as a
bound for $B_{j,c}$. The left-hand side of Eq. (\ref{eq:sumK}) can be written as
\be
\label{eq:sumK2}
\lim_{K\to \infty} \sum_{c=1}^K
\frac{1}{c^{1-w}}\sum_{j=1}^c \left(B_{j,c}\right)_\mu^\nu
\sum_{|d|=K+1}^\infty \frac{e((j+\delta_\nu)\frac{d}{c})}{(c\tau + d)}.
\ee
Ref. \cite{rademacher:1939} gives estimates for the sum over $d$ which
continue to hold for the generalization after minor modifications. We
find that in case $(j+\delta_\nu)/c \in \mathbb{Z}$ for some $j$, the
sum over $d$ has an upperbound given by $\mathcal{O}\left(\frac{c
  \log(K)}{K}\right )$, otherwise the upperbound is
$\mathcal{O}\left(K^{-1}\right )$. The estimates for
Eq. (\ref{eq:sumK}) become respectively, $\lim_{K\to
  \infty}\mathcal{O}\left(K^{w-\epsilon}\log(K)\right )$ and
$\lim_{K\to \infty}\mathcal{O}\left(K^{w-\epsilon}\right )$, which are
indeed zero for ($w<0 $, $\epsilon=0$) and ($w=0$, $\epsilon>0$). We
therefore have shown that Eq. (\ref{eq:sumN}) is equal to 
\be
\lim_{K\to \infty}\sum_{c=1}^K  \sum_{{|d|\leq K \atop
    (c,d)=1}}M^{-1}(\gamma)_\mu^\nu\frac{e((n-\Delta_\nu)\frac{a}{c})}{c^{1-w}(c\tau+d)},
\ee
for the cases which are relevant to us.

The sum over $k$ in Eq. (\ref{eq:subslipschitz}) is equal to an exponent minus the first terms of
the Fourier expansion: $\sum_{k=0}^\infty
\frac{z^{k+1-w}}{\Gamma(k+2-w)}=e^{z}-\sum_{k=0}^{|w|}z^k/k!$, when
$w$ is a negative integer. We recognize the regularization of Eq.
(\ref{eq:regintweight}). However we want to obtain a closed form for
general non-positive weight. This can be obtained using the equality
\begin{eqnarray} \label{hfunction} h(z)=\sum_{k=0}^\infty
\frac{z^{k+1-w}}{\Gamma(k+2-w)}&=&e^z\left(1-\frac{1}{\Gamma(1-w)}\int_z^\infty
e^{-t}t^{-w}dt\right) \\
&=& \frac{e^z}{\Gamma(1-w)} \int_0^z e^{-t}
t^{-w} dt ,\nonumber
\end{eqnarray}
 which is valid for general $w< 1$. One can establish Eq. (\ref{hfunction})
 by developing the second integral expression in series using
 successive integration by parts, or by considering the differential
 equation satisfied by $h(z)$.

We define $R(z)=e^{-z}h(z)$. Inserting this and the equal upperbound
for $c$ and $d$ in Eq. (\ref{eq:subslipschitz}), we obtain
\begin{eqnarray}
f^+_\mu(\tau)&=&\frac{1}{2}F_\mu(\Delta_\mu)+ \sum_{n-\Delta_\nu<0}  \lim_{K\to
  \infty} \sum_{c=1}^{K} \sum_{{|d| \leq K} \atop (c,d)=1}
 \frac{M^{-1}(\gamma)_\mu^\nu F_\nu(n)}{(c\tau+d)^w} \\
&\times& e\left((n-\Delta_\nu)\gamma(\tau)\right)R(x),\nonumber
\end{eqnarray}

\noi where $x=\frac{2\pi i |n-\Delta_\nu|}{c(c\tau+d)}$. The summand
is invariant under $\gamma\to -\gamma$ or equivalently $(c,d)\to
(-c,-d)$. We can extend therefore the sum over $c$ to $0<|c|\leq K$,
and divide by two. The polar part can be included by extending the
sum with $c=0$. Note that $\gcd(0,d)=|d|$, thus $c=0$ adds
$(c,d)=(0,1)$ and $(c,d)=(0,-1)$ to the sum, which works out nicely
with the overall factor of $\frac{1}{2}$. We obtain finally
\begin{eqnarray}
\label{eq:fareytail}
f_\mu(\tau)&=&\frac{1}{2}F_\mu(\Delta_\mu)+ \frac{1}{2}\sum_{n-\Delta_\nu<0}  \lim_{K\to
  \infty}\sum_{\gamma \in (\Gamma_\infty\backslash \Gamma)_K}
j(\gamma,\tau)^{-w} M^{-1}(\gamma)_\mu^\nu F_\nu(n)\\
&&\times e\left((n-\Delta_\nu)\gamma(\tau)\right) R(x), \nonumber
\end{eqnarray}

\noi where we have defined $\sum_{|c|\leq K} \sum_{{|d| \leq K} \atop
  (c,d)=1}=\sum_{\gamma \in (\Gamma_\infty\backslash \Gamma)_K}$.

\subsection{Period functions and their transformation properties}
\label{sec:period}
This subsection reviews relevant properties of period functions. These
properties are necessary for the derivation of the transformation
properties of $f_\mu(\tau)$ in subsection \ref{sec:transformprop}. For
simplicity of exposition we discuss the case of scalar modular
forms. Using the notation of Section \ref{sec:anomalies}, the
discussion generalizes easily to the vector-valued case.

We start with the period function of a cusp form $G(z)$ transforming
as $G(\gamma(z))=M^{-1}(\gamma)(cz+d)^{2-w}G(z)$ under $\gamma\in
\Gamma$.  The period function of $G(z)$, $p(\tau, \bar y, \overline{G})$ is defined
by \be p(\tau,\bar y,\overline{G})=\frac{1}{\Gamma(1-w)}\int_{\bar
y}^{-i\infty}\overline{ G(  z)} (\bar z-\tau)^{-w}d\bar z, \qquad y
\in \mathcal{H}\cup \mathbb{Q} \cup i\infty. \ee

\noi Note that in case $-w \in \mathbb{N}$, this expression is a
  polynomial in $\tau$. Also note that the expression $p(\tau, \bar y, \overline{G})$
  makes sense for any function $G(z)$ that decays sufficiently rapidly at
  infinity, e.g. $G(x+i \rho) \sim_{\rho\to + \infty} const. \rho^\alpha e^{-A \rho}$ for $A>0$
  will suffice.  The constituents of the integrand satisfy simple
transformation properties: $\gamma(\bar z)-\gamma(\tau)=\frac{\bar z
  -\tau}{j(\gamma,\bar z) j(\gamma,\tau)}$ and
$d\gamma(z)=\frac{dz}{j(\gamma,z)^2}$. Using these equations we
obtain for $p(\gamma(\tau),\gamma(\bar y),\overline{G(z)})$ the transformation rule
\be \label{eq:transformp}p(\gamma(\tau),\gamma(\bar y),
\overline{G})=j(\gamma,\tau)^w M(\gamma)\left[p(\tau,\bar y,
\overline{G})-p(\tau,\gamma^{-1}(\infty), \overline{G}) \right], \ee
where we have used the fact that $M(\gamma)$ is unitary.

If we choose a constant $\delta >0$ we can try to construct a cusp
form   $G^{(\delta)}(z)$ of weight $2-w$ by forming the  Poincar\'e
series
  \be \label{eq:cuspform} G^{(\delta)}(z)=\frac{1}{2}\sum_{\gamma\in \Gamma_\infty\backslash \Gamma}
  \frac{ M(\gamma)(-2\pi
i\delta)^{1-w} e(\delta \gamma(z))}{j(\gamma,z)^{2-w}}:=\frac{1}{2}\sum_{\gamma\in \Gamma_\infty\backslash \Gamma}
  g_\gamma^{(\delta)}(z),
\ee

\noi where we defined $g_\gamma^{(\delta)}(z)$ by the second
equality. The prefactor is chosen for later convenience. We will
sometimes drop the superscript $\delta$ when the context is clear.
For $w<0$ the series is convergent, although it might vanish.

The period functions are relevant for our discussion of the
fareytail expansions as explained in appendix
\ref{sec:transformprop} and section \ref{sec:anomalies}. In those
discussions we make use of the function   $t_\gamma(\tau)$ defined
by
 \be
t_\gamma(\tau):=p(\tau,\gamma^{-1}(i\infty),\overline{g_\gamma}).
\ee

\noi Using the above identities and Eq. (\ref{eq:jcocycle}) one can
check that  $t_\gamma(\tau)$ satisfies the transformation rule with
$\tilde \gamma \in \Gamma$ \be \label{eq:transformt} t_\gamma(\tilde
\gamma(\tau))=j(\tilde\gamma,\tau)^w M(\tilde \gamma)
\left[t_{\gamma\tilde \gamma}(\tau)-p(\tau,\tilde
  \gamma^{-1}(i\infty),\overline{g_{\gamma
    \tilde \gamma}}) \right],
\ee

\noi Note that  $t_\gamma(\tau)$ can be rewritten as
\begin{eqnarray}
\label{eq:tintegral} t_\gamma(\tau)&=&\frac{-1}{\Gamma(1-w)}
j(\gamma,\tau)^{-w} M^{-1}(\gamma)e(-\delta\gamma(\tau))
\int_x^\infty e^{-z}z^{-w}dz,
\end{eqnarray}

\noi with $x=\frac{2\pi i \delta}{c j(\gamma,\tau)}$ where $c$
is the $21$ matrix element of $\gamma$. The steps involved are first
a transformation of $\bar z$ to $\gamma^{-1}(\bar z)$, then
rewriting of the integrand using its modular properties and at last
another redefinition of $\bar z$.

\subsection{Transformation properties of the fareytail}
\label{sec:transformprop} We will deduce the transformation
properties of $f_\mu(\tau)$ from the expression given in Eq.
(\ref{eq:fareytail}). Many intermediate steps are given without
rigorous proofs, these can be found in Ref. \cite{niebur1974}. We
discuss the case of scalar modular forms; at the end we simply state
the straightforward generalization to vector-valued modular forms.
The discussion reverses the logic of Section \ref{sec:anomalies}.

We study first the transformation properties of a (scalar) modular
form with a single polar term $q^{-\delta}$ ($\delta>0$) for a clear
exposition. Eventually we will deduce the transformation law for
general $f_\mu(\tau)$. We define the function
$s_\gamma(\tau)=j(\gamma,\tau)^{-w}
M^{-1}(\gamma)e(-\delta\gamma(\tau)) $ and use $t_\gamma(\tau)$ as
in Eq. (\ref{eq:tintegral}). Eq. (\ref{eq:fareytail}) is in this
case given by

\be
f^{(-\delta)}(\tau)=\frac{1}{2}F(\delta)+\frac{1}{2} \lim_{K\to
  \infty}\sum_{\gamma \in (\Gamma_\infty\backslash \Gamma)_K}s_\gamma(\tau)+t_\gamma(\tau).
\ee

\noi $s_\gamma(\tau)$ satisfies $s_\gamma(\tilde
\gamma(\tau))=j(\tilde \gamma,\tau)^w M(\tilde \gamma)
s_{\gamma\tilde \gamma}(\tau)$. We obtain with Eq.
(\ref{eq:transformt})
\begin{eqnarray}
\label{eq:fdelta}
f^{(-\delta)}(\tilde \gamma(\tau))&=&\frac{1}{2}F(\delta) \\
&+&\frac{1}{2} M(\tilde \gamma)(\tilde c \tau +\tilde
d)^w \lim_{K\to
  \infty}\sum_{\gamma \in (\Gamma_\infty\backslash \Gamma)_K}s_{\gamma\tilde
  \gamma}(\tau)+t_{\gamma \tilde \gamma}(\tau)-p(\tau,\tilde
  \gamma^{-1}(-i\infty), \overline{g_{\gamma
    \tilde \gamma}}). \nonumber
\end{eqnarray}

\noi The invariance under $T=\gamma_1$ is obvious from the Fourier
  expansion and Eq. (\ref{eq:subslipschitz}). We therefore only need to
  check the invariance under the other generator of $\Gamma$, $S=\left(\begin{array}{cc}0 & -1 \\ 1 &
  0 \end{array}\right)$. $(\Gamma_\infty\backslash \Gamma)_K$ is
  however left invariant under right multiplication of $S$. Therefore,
  $\sum_{\gamma \in (\Gamma_\infty\backslash \Gamma)_K}s_{\gamma S}(\tau)+ t_{\gamma S}(\tau)=\sum_{\gamma \in (\Gamma_\infty\backslash
  \Gamma)_K}s_{\gamma}(\tau)+ t_{\gamma}(\tau)$ holds.

 The anomalous terms compared to the usual transformation rule of modular forms are the
  constant term $\frac{1}{2}F(\delta)$ and
the subtraction of period integrals. A careful study of the limit
$K\to \infty$ and the period integrals is needed. Lemma 4.4 of
Ref. \cite{niebur1974} shows that for  $y\in \mathcal{H}$
\be \lim_{K\to \infty}
\sum_{\gamma \in (\Gamma_\infty\backslash \Gamma)_K}
  p(\tau,\bar y,\overline{g_{\gamma}^{(\delta)}})= p(\tau,\bar y,\overline{G^{(\delta)}})-F(\delta),
\ee

\noi thus the limit $K\to \infty$ and the integral do not
commute. This comes about as follows. Calculation of the Fourier
coefficients of $G^{(\delta)}$ gives an error term by the Lipschitz
summation formula. This error term tends to zero, however the period
integral over the error does not vanish and provides us with the
offset.

In Eq. (\ref{eq:fdelta}), we however have $y\not\in \mathcal{H}$ but
$y=\tilde \gamma^{-1}(i\infty)\in \mathbb{Q}$. In this case we obtain
with Corollary 4.5 of Ref. \cite{niebur1974}
\begin{eqnarray}
\lim_{K\to \infty} \sum_{\gamma \in (\Gamma_\infty\backslash \Gamma)_K}
  p(\tau, \tilde \gamma^{-1}(i\infty) ,\overline{g_{\gamma}^{(\delta)}})&=&
  p(\tau,\tilde \gamma^{-1}(i\infty),\overline{G^{(\delta)}}) \\
&&+F(\delta)\left(M^{-1}(\tilde
  \gamma)(\tilde c\tau + \tilde d)^{-w}-1 \right). \nonumber
\end{eqnarray}

\noi Inserting this result in Eq. (\ref{eq:fdelta}) we find the
transformation of $f^{(-\delta)}(\tau)$ under $\gamma$
\be \label{eq:shifttrnsm}
f^{(-\delta)}(\gamma(\tau))= j(\gamma,\tau)^w
M(\gamma) \left[
f^{(-\delta)}(\tau)_\delta-p(\tau,\gamma^{-1}(i\infty),\overline{G^{(\delta)}})\right].
\ee

\noi Note that in special cases $G$ is zero. This is for example the case
for $\delta \in \mathbb{N}$ and $w=0,-2,-4,-6,-8$ and $-12$
\cite{knopp:1990}. A cusp form with weight $12=2-w$ of $\Gamma$
exists, which explains that in case $w=-10$, we will find a
transformation with a non-zero shift.

Extending the above to the case of vector-valued modular forms with multiple
polar terms is straightforward. The period function should vanish of
course in this case. For a general choice of $\Delta_\mu$ and polar
$F_\mu(n)$, we obtain the transformation
\be
\label{eq:transformvectorform}
f_\mu(\gamma(\tau))= (c \tau + d)^w M(\gamma)_\mu^\nu  \left[ f_\nu(\tau)-p(\tau,\gamma^{-1}(-i\infty),\overline{G_\nu})\right].
\ee

\noi with
\be
G_\mu(z)=\frac{1}{2}\sum_{n-\Delta_\nu<0} \sum_{\gamma \in \Gamma_\infty\backslash
  \Gamma} \overline{M^{-1}(\gamma)^\nu_\mu} \frac{(2\pi i
(n-\Delta_\nu))^{1-w}F_\nu(n) e(|n-\Delta_\nu|\gamma(z))}{(cz+d)^{2-w}}.
\ee

\section{Lipschitz Summation Formula}
\label{sec:lipschitz}\setcounter{equation}{0}
A crucial ingredient for the derivation in
appendix \ref{sec:derivation} is the
Lipschitz summation formula for general $p \geq 1$ \cite{niebur1974}. Let $\tau \in
\mathcal{H}$, $N\in \mathbb{N}$, $0\leq \alpha < 1$, then
 \be
\label{eq:lipschitz} \sum_{l=-N}^N \frac{e(-l\alpha)}{(\tau +
l)^p}=\frac{(-2\pi i)^p}{\Gamma(p)}\sum_{m=0}^\infty
(m+\alpha)^{p-1}q^{m+\alpha} + E(\tau,p,Q), 
\ee

\noi where $Q= N+\frac{1}{2}$ and  $E(\tau,p, Q)$ is an error term
and given by
\be
E(\tau,p,Q)=(iQ)^{1-p}\int_{-\infty}^\infty
\frac{h(x-i)-h(x+i)}{1+\exp(2\pi x
  Q)}dx, \qquad h(x)=\frac{\exp(2\pi x Q \alpha)}{(x+\frac{\tau}{iQ})^{p}},
\ee

\noi  The error tends to 0 for $Q \to \infty$,
except for the case $p=1$, $\alpha=0$; then we obtain $\lim_{Q\to\infty}
E(\tau,1,Q)=\pi i$. The
case $p=1$, $\alpha=0$ gives the two well known infinite sums for
$\cot \pi \tau$
\be
\frac{1}{\tau}+\sum_{l=1}^\infty\left(\frac{1}{\tau-l}+\frac{1}{\tau+l}\right)=\pi \cot \pi \tau=\pi i - 2\pi i \sum_{m=0}^\infty q^m,
\ee

\noi which can be proved by using $\sin \pi \tau=\pi \tau
\prod_{n=1}^\infty (1-\tau^2/n^2)$.

The proof of Eq. (\ref{eq:lipschitz}) uses the function
$f(z)=e((z+\tau)\alpha)/(iz)^p(e(z+\tau)-1)$. This function has
poles at $z=-\tau-l$, $l\in \mathbb{Z}$ with residue $(2\pi i)^{-1}
e(-l\alpha)/(-i\tau-il)^p$. The right hand side is obtained by
integrating along the boundary of the rectangle $-\re(\tau)\pm Q \pm
iM$, which is slit along the positive imaginary axis to avoid a
branch cut of $(iz)^p$. The main contribution to the integral comes
from this part of the contour. It can be calculated using the Hankel
contour integral $\frac{1}{\Gamma(p)}=\frac{1}{2\pi i} \int_{\CC}e^t
t^{-p} dt$, where $\CC$ is the contour which begins at $-\infty-i
0^+$, circles the origin in the counterclockwise direction and ends
at $-\infty + i 0^+$.  The horizontal sides do not contribute when
$M\to \infty$, the error is accordingly calculated by the integral
along the vertical segments.

\section{Details on Multiplier Systems}
\label{sec:multisystems}\setcounter{equation}{0}

We remark that  consistency of Eq. (\ref{eq:modtransform}) requires
$M(\gamma)$ to satisfy
\be \label{eq:g1g2} M(\gamma_1)_\mu^\rho M(\gamma_2)_\rho^{\nu} =
c_w(\gamma_1,\gamma_2) M(\gamma_1 \gamma_2)_\mu^\nu, \ee
where
\be \label{eq:sl2cocyc} c_w(\gamma_1,\gamma_2) :=
\frac{j(\gamma_1\gamma_2,\tau)^w}{j(\gamma_1,\gamma_2\tau)^w
j(\gamma_2,\tau)^w}. \ee
Using the identity
\be \label{eq:jcocycle} j(\gamma_1\gamma_2,\tau) =
j(\gamma_1,\gamma_2 \tau) j(\gamma_2,\tau), \ee
we see that the right hand side of Eq. (\ref{eq:sl2cocyc}) is a phase.
On the other hand, it is locally analytic in $\tau$, and hence it
does not depend on  $\tau$. Indeed $c_w(\gamma_1,\gamma_2)$ is a
cocycle on $\Gamma$.  Then the cocycle is most easily evaluated by
taking $\tau = i \Lambda, \Lambda \to + \infty$. Define
$\epsilon(\gamma)=\pm 1$ by
\be  \epsilon(\gamma):= \begin{cases} {\rm sign}(c), & c \not=0, \\
{\rm sign}(d), & c=0. \\ \end{cases} \ee
Then we have with $\epsilon_i=\epsilon(\gamma_i)$
\be \label{eq:evalcocyc} c_w(\gamma_1,\gamma_2)=
  \exp\left[ \frac{i \pi}{2} w (\epsilon_1\epsilon_2
\epsilon_3 - \epsilon_1 - \epsilon_2 + \epsilon_3) \right]. \ee
where  $\gamma_3=\gamma_1\gamma_2$. This expression takes values $1,
e^{\pm 2\pi i w} $.

Note that

\begin{enumerate}

\item $c_w$ is symmetric and $c_w(1,\gamma)=c_w(\gamma,1)=1$,

\item $M(-\gamma)_\mu^\nu = e^{i \pi w \epsilon(\gamma)}
M(\gamma)_\mu^\nu$,

\item It is perfectly possible to have $(\epsilon_1,\epsilon_2,\epsilon_3)
= (-1,-1,+1)$. For example, take
  $$\gamma_1=\left(\begin{array}{cc} N+1 & N\\ -N & 1-N
\end{array}\right), \qquad \gamma_2=\left(\begin{array}{cc} 1 & 0\\ -N & 1
\end{array}\right),$$
 with  $N>2$, thus realizing
$c_w(\gamma_1,\gamma_2) = e^{2\pi i w }$.

\end{enumerate}

In applications to the elliptic genus it is possible to describe the
multiplier system explicitly. In the case of the $(2,2)$ elliptic genus, in order to have a basis
of linearly independent functions we should make a unitary
transformation to the even and odd level $m$ theta functions and
correspondingly define $f_\mu$ by expanding with respect to the even
level $m$ theta functions, defined by
\be \theta_{\mu,m}^+(\tau,z) := \begin{cases} \theta_{0,m}(\tau,z), &
\mu=0, \\
\frac{1}{\sqrt{2}}(\theta_{\mu,m}(\tau,z) + \theta_{-\mu,m}(\tau,z)),
& 1\leq \mu \leq m-1, \\
\theta_{m,m}(\tau,z), & \mu=m.  \\ \end{cases} \ee
and defining $\phi(\tau,z):=\sum_{\mu=0}^{m} h_\mu^+(\tau)
\theta_{\mu,m}^+(\tau,z)$.
 Taking $S= \begin{pmatrix} 0 & 1 \\ -1 & 0 \\
\end{pmatrix}$ we find
\be M(S) = e^{-i \pi/4} \begin{pmatrix} S_{00}=\frac{1}{\sqrt{2m}} &
S_{0,\mu}=\frac{1}{\sqrt{m}} &
S_{0,m}=\frac{1}{\sqrt{2m}} \\
S_{\mu,0}=\frac{1}{\sqrt{m}} & S_{\mu\nu}=\sqrt{\frac{2}{m} }
\cos\left(2\pi \frac{\mu\nu}{2m}\right) &
S_{\mu,m}=\frac{(-1)^\mu}{\sqrt{m}} \\
S_{m,0}=\frac{1}{\sqrt{2m}} & S_{m,\mu}=\frac{(-1)^\mu}{\sqrt{m}} &
S_{m,m}=\frac{(-1)^m}{\sqrt{2m}} \\ \end{pmatrix}. \ee
where $1\leq \mu,\nu\leq m-1$ in the above matrix. Of course, we
also have
\be M(T)_\mu^\nu = e\left(-\frac{\mu^2}{4m}\right) \delta_\mu^\nu. \ee
Together these generate the multiplier system.

\bibliographystyle{JHEP}

\providecommand{\href}[2]{#2}\begingroup\raggedright\endgroup

\end{document}